\documentclass[a4paper, 11pt]{article}
\usepackage{amsmath,amssymb,natbib}
\usepackage{graphicx,color,soul}
\usepackage[usenames,dvipsnames]{xcolor}
\usepackage[displaymath]{lineno}
\usepackage[hmargin=2.5cm, vmargin=3cm]{geometry}

\title{Dilatancy toughening of shear cracks and implications for slow rupture propagation}

\author{Nicolas Brantut\\Department of Earth Sciences\\ University College London, London, UK.}

\date{\ }

\begin{document}

\maketitle


\begin{abstract}
  Dilatancy associated with fault slip produces a transient pore pressure drop which increases frictional strength. This effect is analysed in a steadily propagating rupture model that includes frictional weakening, slip-dependent fault dilation and fluid flow. Dilatancy is shown to increase the stress intensity factor required to propagate the rupture tip. With increasing rupture speed, an undrained (strengthened) region develops near the tip and extends beyond the frictionally weakened zone. Away from the undrained region, pore fluid diffusion gradually recharges the fault and strength returns to the drained, weakened value. For sufficiently large rupture dimensions, the dilation-induced strength increase near the tip is equivalent to an increase in toughness that is proportional to the square root of the rupture speed. In general, dilation has the effect of increasing the stress required for rupture growth by decreasing the stress drop along the crack. Thermal pressurisation has the potential to compensate for the dilatant strengthening effect, at the expense of an increased heating rate, which might lead to premature frictional melting. Using reasonable laboratory parameters, the dilatancy-toughening effect leads to rupture dynamics that is quantitatively consistent with the dynamics of observed slow slip events in subduction zones.
\end{abstract}


\section{Introduction}

Dilatancy is a well documented process in intact, healed or overconsolidated rocks: due to shear deformation, microcavities (cracks, grain junctions) open, which induces an overall increase in porosity \citep[][Section 5.3]{paterson05}. In fluid saturated rocks, this increase in porosity has the potential to induce fluid pressure drops if the dilatant region is insufficiently drained, producing an increase in effective stress and thus strengthening the material. This phenomenon is called dilatancy hardening.

The importance of dilatancy hardening in fault mechanics has long been recognised in laboratory experiments and in theoretical rupture models. One of the first experimental evidence of dilatancy hardening in crustal rocks was obtained by \citet{brace68} in diabase and granite, who showed that failure strength was increasing with increasing deformation rate as the deformation conditions became more undrained. Similar observations have been made on a range of low-porosity rocks \citep[e.g.,][]{rutter72,chiu83,duda13}. \citet{martin80} further demonstrated that rock failure could be stabilised due to dilatancy, by considerably slowing down the deformation leading to strain localisation and fault slip. New laboratory results by \citet{aben21} have confirmed the direct stabilisation of ruptures due to shear-induced dilation. Such work was focused on initially intact rock, where dilatancy is occurring in the bulk as well as within the incipient fault zone. In preexisting fault zones, dilatancy can be due to overriding asperities (along bare rock surfaces) and granular rearrangements (if gouge layers are present), leading to net fault zone opening during slip. Such behaviour has been thoroughly documented in artificial fault gouges, for which dilatancy can be related to frictional state evolution \citep[e.g.,][]{marone90,sleep06,samuelson09}. Recent laboratory studies have shown that fault zone dilatancy (and compaction) exerts a first-order control on fluid pressure \citep{lockner94,faulkner18,brantut20} and slip dynamics \citep{proctor20,aben21}.



From a theoretical point of view, the role of dilatancy hardening in the propagation of shear rupture was initially investigated by \citet{rice73}, who showed that dilatancy tends to increase the fracture energy necessary to drive rupture propagation. \citet{rice79} demonstrated that dilatant hardening leads to a period of stable, quasi-static deformation prior to shear rupture instability. In a one dimensional (spring-slider) fault zone model in the context of slip-dependent strength and dilation, \citet{rudnicki88} showed that dilatancy could prevent unstable fault motion. A similar stabilisation effect was found by \citet{segall95} for faults governed by rate and state dependent friction and dilation. For spatially extended fault slip, the competition between dilation, fluid flow and frictional weakening has been shown to produce slowly propagating slip events, in a manner and parameter range consistent with geophysical observations of slow slip in subduction zones \citep{liu10,segall10,liu13}. Dilatancy has also been shown to promote quasi-static rupture propagation in coupled thermo-hydro-mechanical models where the weakening effect of thermal pressurisation of pore fluid is limited by the strengthening effect of fault zone dilation \citep[e.g.,][]{suzuki07,suzuki08,suzuki09}.

Most of the aforementioned rupture models including the effect of fault zone dilation have relied on numerical solutions for either the quasi-static \citep{segall10} or dynamic \citep{suzuki08} equation of motion, effectively solving the fully coupled, nonlinear fracture problem without relying on specific assumptions regarding rupture process zones and weakening behaviour. \citet{liu10} developed an energy balance approach based on fracture mechanics to determine analytical estimates for slip rate during slow slip transients, which compared well with their numerical results. The success of this approach is further justified by recent work by \citet{barras20,garagash21} which demonstrated, in the context of rate and state friction laws, that nonlinear solutions to rupture problems could be recast in the framework of linear elastic fracture mechanics, where rupture propagation is governed by a Griffith-type criterion $G=G_\mathrm{c}$, where $G$ is the energy-release rate and $G_\mathrm{c}$ is a fracture energy that depends on rupture speed and parameters of the nonlinear friction law. Such a connection between nonlinear problems and linear elastic fracture mechanics equivalents was already highlighted by \citet{ampuero08, hawthorne13} for rate and state friction law. Estimates of fracture energy were also given by \citet[][\S 78, 79 and 80]{segall10} for the coupled, quasi-static, rate and state friction and dilatancy-diffusion problem. While those estimates are very useful to analyse simulation results \emph{a posteriori}, they are expressed as a function of slip rate, which is \emph{a priori} unknown in fracture problems. In order to use fracture energy in a Griffith energy balance $G=G_\mathrm{c}$, or, equivalently, in a balance between stress intensity factor $K$ and toughness $K_\mathrm{c}$, and arrive at elementary predictions for rupture tip dynamics (a time and space history of the crack tip position), fracture energy or toughness would need to be expressed as a function of rupture speed instead of fault slip rate \citep[as done recently by][]{garagash21}.

In addition to its potential impact on stability and fracture energy, one other key aspect of dilatancy in the context of rapid fault slip is its potential to counteract thermal pressurisation of pore fluids and to accelerate heat production. This effect was explored extensively by \citet{garagash03a}, and some solutions of the coupled thermo-hydro-mechanical problem were exposed by \citet{segall12}. For dilatancy occurring much more rapidly than shear heating, it amounts to resetting the pore pressure inside the fault at the onset of slip \citep[][\S 50]{rice06}, which increases the potential maximum temperature achieved during slip. This phenomenon plays a major role in determining the onset of frictional melting in low porosity, consolidated rocks \citep{brantut18c}; in granite, dilatancy can indeed be so large as to rapidly decrease pore pressure down to vapor pressure during faulting and slip \citep{brantut20}. Recent laboratory measurements of fault zone dilatancy and hydraulic properties during dynamic rupture events \citep{brantut20} highlight the need for a comprehensive assessment of the effect of dilatancy on rupture dynamics.

The present study focuses attention on the effects of fault zone dilatancy on the weakening and thermal response of the fault during dynamic slip, and aims to find a simple description of those effects in terms of fracture energy, to be used in a Griffith-type energy balance. Based on recent laboratory results \citep{brantut20,aben21}, a simple slip-dependent weakening and porosity change model is used (similar to the approach of \citet{rudnicki88}). Although more advanced parameterisations of weakening and dilation based on rate-and-state descriptions of friction  have been determined experimentally at slow slip rates and used in theoretical investigations \citep[e.g.,][]{marone90,segall95,sleep06,samuelson09}, a direct dependence of friction and dilation on slip is a limit case of rate-and-state descriptions in response to sudden changes in slip rate, which is particularly relevant to rapid propagation of shear ruptures. A steadily propagating dynamic crack tip model is employed to arrive at semi-analytical results for shear stress, pore pressure and temperature evolution along the propagating rupture. Keeping with the eventual goal of finding a simple, usable form of fracture energy to use in a Griffith energy balance, a particular attention is paid to the contribution of fault zone dilation to fracture energy as a function of rupture propagation speed. This analysis leads to a refined description of the various contributions to fracture energy and energy release rate of different weakening processes, including ``near-tip'' mechanisms such as intrinsic frictional weakening or dilatancy, which contribute to $G_\mathrm{c}$, and ``far-tip'' mechanisms like pore fluid diffusion or thermal pressurisation (or decomposition, or melting), which are expected to contribute to $G$.

\section{Model}

\begin{figure}
  \centering
  \includegraphics{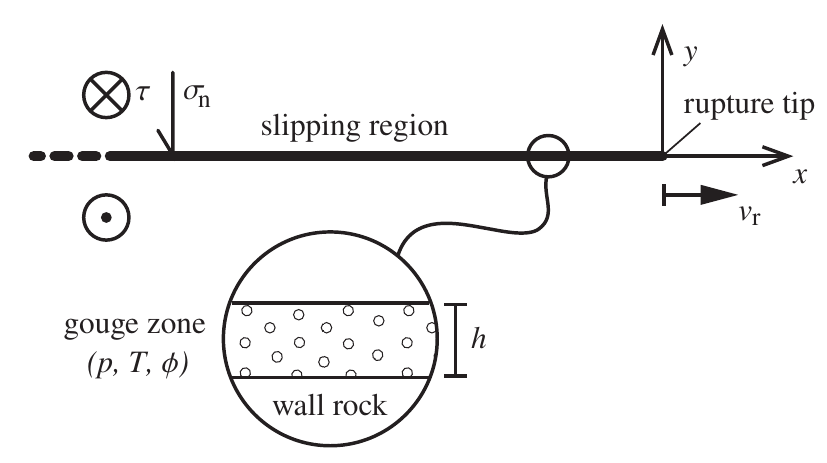}
  \caption{Schematic of the model setup, with dynamically propagating rupture at speed $v_\mathrm{r}$ along the $x$ axis under anti-plane geometry. The internal structure of the fault is considered to be a porous material (gouge) of finite width (in the $y$ direction), undergoing shear and dilation.}
  \label{fig:sketch}
\end{figure}

\begin{figure}
  \centering
  \includegraphics{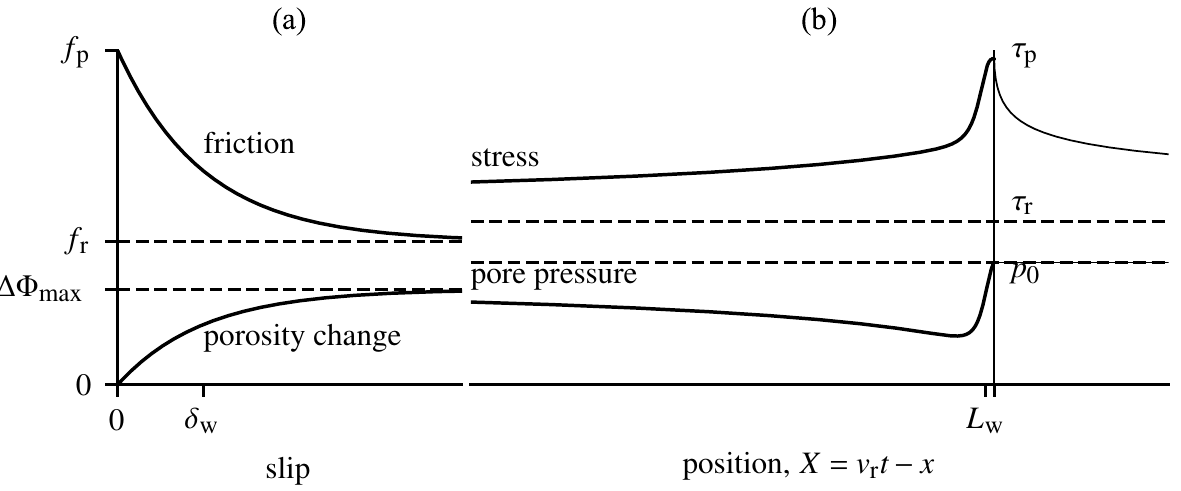}
  \caption{(a) Constitutive laws for friction and porosity evolution with slip (Equations \ref{eq:f} and \ref{eq:Dphi}); (b) Propagating crack model, with stress change due to both friction and pore pressure changes (Equation \ref{eq:tauf}). The tip is mostly undrained, and recharge occurs at large distances.}
  \label{fig:setup}
 \end{figure}



\subsection{Elastodynamics}

Our main focus is to determine the relative contributions of slip-weakening, dilatancy and thermal pressurisation in the energy balance of dynamic ruptures. A fully dynamic approach would require extensive numerical simulations, imposing somewhat arbitrary choices of initial and boundary conditions. For the purpose of making estimates of fracture energy and dissipation consistent with elastodynamics, it is useful to consider only the tip region of a dynamically propagating, steady-state, semi-infinite shear rupture \citep{viesca15}. The steady-state assumption corresponds to an assumption of constant rupture speed $v_\mathrm{r}$; it is valid when rupture speed varies only smoothly as a function of time, over distances (or timescales) much larger than the characteristic dimensions of the non-linear tip region. In a two dimensional configuration where the fault lies along the $x$ axis (Figure \ref{fig:sketch}), the shear stress $\tau$ at position $X=v_\mathrm{r}t-x$ (in a reference frame moving with the rupture tip, see Figure \ref{fig:setup}b) is related to the slip rate distribution $V(X)$ as
\begin{linenomath}
\begin{equation} \label{eq:elastodyn}
  \tau(X) = \tau_\mathrm{b} + \frac{\bar{\mu}}{2\pi v_\mathrm{r}}\int_0^\infty \frac{V(\xi)}{\xi-X}d\xi,\quad\text{(steady-state)}
\end{equation}
\end{linenomath}
where $\bar{\mu}$ is an apparent elastic modulus that depends on loading mode and rupture velocity. For semi-infinite ruptures, stress drop is neglected compared to strength variations and $\tau_\mathrm{b}$ is set equal to the far-field residual shear strength on the fault. For simplicity, we only consider mode III (anti-plane) loading here; in this case $\bar{\mu} = F(v_\mathrm{r}/c_\mathrm{s})\mu$ where $\mu$ is the shear modulus of the fault walls, $c_\mathrm{s}$ is the shear wave speed, and $F(z)=\sqrt{1-z^2}$ \citep[e.g.,][]{rice80}. In the moving reference frame at constant speed $v_\mathrm{r}$, the partial time derivatives are effectively derivatives with respect to coordinate $X$, $\partial/\partial t = (1/v_\mathrm{r})\partial/\partial X$, so that the dependent variables required to solve the dynamic problem are functions of $X$ only.


\subsection{Shear strength}

In the slipping part of the fault, the shear stress in Equation \ref{eq:elastodyn} must equal the shear strength of the fault. We assume that the fault is of finite width, filled with a porous material (Figure \ref{fig:sketch}). It is initially saturated with a pore fluid, considered chemically inert at the timescale of the phenomena of interest here, at a uniform pressure $p=p_0$. The shear resistance to sliding, $\tau_\mathrm{f}$, is given by Terzaghi's effective stress principle,
\begin{linenomath}
\begin{equation} \label{eq:tauf}
  \tau_\mathrm{f} = f\times\sigma',
\end{equation}
\end{linenomath}
where $\sigma'=\sigma_\mathrm{n}-p$ is the effective normal stress, $\sigma_\mathrm{n}$ is the applied normal stress, considered constant, and $f$ is the friction coefficient.

The fault zone is assumed to be made of a consolidated material, for instance a healed cataclasite or fault gouge, so that shear strain produces significant (1) ``direct'' weakening of the fault by a decrease of the friction coefficient from a peak $f_\mathrm{p}$ to a residual value $f_\mathrm{r}$, and (2)
dilation of the fault, by way of microcrack opening and overriding of micro- and meso-scale asperities \citep[e.g.][]{barton76}.

\subsection{Slip weakening behaviour}

The former ``direct weakening'' effect has been thoroughly documented experimentally in consolidated or healed rocks \citep[e.g.,][]{karner97,nakatani04}, and can be thought of as an irreversible decrease in cohesion; while the physical processes leading to the progressive degradation of strength with ongoing slip remain difficult to constrain in detail, it suffices for the present analysis to capture such a weakening effect as a simple progressive slip-weakening behaviour for the friction coefficient (see a similar analysis in \citet{rudnicki88}; Figure \ref{fig:setup}a),
\begin{linenomath}
\begin{equation} \label{eq:f}
  f(\delta) = f_\mathrm{r} + (f_\mathrm{p}-f_\mathrm{r})e^{-\delta/\delta_\mathrm{w}},
\end{equation}
\end{linenomath}
where $\delta$ denotes the total slip across the fault, and $\delta_\mathrm{w}$ denotes a characteristic slip-weakening distance.

In the context of fast slip (with no subsequent healing), the frictional weakening as phenomenologically described by \eqref{eq:f} is not fundamentally different from more elaborate formulations based on rate-and-state friction laws including rapid weakening at high slip rate, for instance due to flash heating \citep{noda09,brantut11d,viesca15}. For dynamic weakening, peak friction is typically of the order of $1$ (or, $0.6$ to $0.8$ if we strictly follow Byerlee's rule of thumb), and weakened friction can be as low as $0.1$, for instance due to asperity-scale thermal weakening \citep[e.g.,][]{rice06,ditoro11}. The slip weakening distance $\delta_\mathrm{w}$ for dynamic weakening could be as small as a few 10s of $\mu$m \citep[e.g.,][]{noda09,brantut11d,viesca15}, but there is still considerable uncertainty in choosing this parameter in the phenomenological description \eqref{eq:f}.

At slower, sub-seismic slip rates, frictional weakening as in \eqref{eq:f} is consistent with commonly used rate-and-state friction laws in rate-weakening materials and in the absence of healing: a sudden increase in slip rate does produce an exponential decay of the friction coefficient from a peak (proportional to the log of the velocity increase) to a residual value (proportional to the new ``state'' of the interface) \citep[][, chap. 2]{scholz02}.

If we focus our attention to slip-weakening behaviour observed in association with decreases in cohesion (e.g., during failure of intact or consolidated rocks), then appropriate values for peak frictional strength could be as high as $1.5$, and the weakened friction coefficient is then of the order of $0.6$ to $0.8$ \citep{byerlee78}. The slip weakening distance $\delta_\mathrm{w}$ typically associated with the failure process is of the order of $0.1$ to $1$~mm \citep[e.g.,][]{wong82,lockner91,ohnaka99,ohnaka03,aben19}.


\subsection{Slip-dependent dilatancy}

Fault zone dilation due to shear deformation is modelled following the approach of \citet{rudnicki88}, and we consider that the inelastic porosity change of the fault material depends directly on slip as (Figure \ref{fig:setup}a)
\begin{linenomath}
\begin{equation} \label{eq:Dphi}
  \Delta\Phi(\delta) = \Delta\Phi_\mathrm{max}(1-e^{-\delta/\delta_\mathrm{D}}),
\end{equation}
\end{linenomath}
where $\Delta\Phi_\mathrm{max}$ is the maximum inelastic porosity change at large slip, and $\delta_\mathrm{D}$ is a characteristic slip distance associated with the change in porosity.

A phenomenological relationship like \eqref{eq:Dphi} is  supported by experimental data in consolidated rocks \citep[e.g.,][]{barton76,teufel81,brantut20,aben21}. In this description, the parameter $\Delta\Phi_\mathrm{max}$ can be thought of as the porosity change achieved when the fault zone material reaches critical state, i.e., the stage at which further shear deformation does not impart any bulk volume change. In the case of dilatancy produced by shear rupture in consolidated rocks, the characteristic slip $\delta_\mathrm{D}$ is likely commensurate to the slip weakening distance $\delta_\mathrm{w}$, since both the weakening and the dilatancy phenomena are driven by the same underlying processes (microcrack opening, linkage and slip). This was the case reported by \citet{brantut20,aben21} in initially intact granite, where $\delta_\mathrm{D}$ of the order of $1$~mm was determined from volume change vs. slip data.

The porosity evolution given by \eqref{eq:Dphi} is also compatible with the rate-and-state description given by \citet{segall95} for faults subject to sudden changes in slip rate. In that framework, $\Delta\Phi_\mathrm{max}$ is given by a dilatancy factor, of the order of $10^{-3}$, multiplied to the log of the velocity step, and the characteristic slip $\delta_\mathrm{D}$ is equal to the characteristic slip for state evolution, of the order of 10s of $\mu$m.

Overall, the phenomenological description \eqref{eq:Dphi} is consistent with both shear failure of consolidated materials and slip across preexisting discontinuities or gouge layers (in the ``no-healing'' limit), provided that the value of parameters $\Delta\Phi_\mathrm{max}$ and $\delta_\mathrm{D}$ are chosen accordingly. Such a versatile description, with a limited number of parameters, is useful to capture the essence of the dilatancy phenomenon.


\subsection{Fluid flow and shear heating}

Due to dilatancy, frictional heating and hydraulic diffusion, the fault zone pore pressure $p$ in \eqref{eq:tauf} evolves as a function of time $t$ during slip. The governing equations for thermal pressurisation have been given numerous times in the literature, and only essential steps are recalled here. All notations directly follow those of \citet{rice06,garagash12}. The conservation of fluid mass leads to the following governing equation for $p$ \citep[e.g.,][]{rice06}:
\begin{linenomath}
\begin{equation}\label{eq:dpdt}
  \frac{\partial p}{\partial t} = \Lambda \frac{\partial \Theta}{\partial t} + \alpha_\mathrm{hy}\frac{\partial^2p}{\partial y^2} - \frac{1}{\beta^*}\frac{\partial n^\mathrm{pl}}{\partial t},
\end{equation}
\end{linenomath}
where $\Lambda$ is the thermal pressurisation coefficient, $\Theta$ is temperature, $\alpha_\mathrm{hy}$ is the hydraulic diffusivity and $\beta^*$ is a hydraulic storage capacity of the fault zone material. The local inelastic change in porosity is denoted $n^\mathrm{pl}$. The coordinate $y$ is oriented perpendicular to the fault plane (Figure \ref{fig:sketch}), and we neglect along-fault fluid diffusion.

The temperature evolution of the fault zone is given by \citep{rice06}
\begin{linenomath}
\begin{equation}\label{eq:dTdt}
  \frac{\partial \Theta}{\partial t} = \frac{\tau \dot{\gamma}}{\rho c} + \alpha_\mathrm{th}\frac{\partial^2\Theta}{\partial y^2},
\end{equation}
\end{linenomath}
where $\dot{\gamma}$ is the shear strain rate, $\rho c$ is the heat capacity and $\alpha_\mathrm{th}$ is the thermal diffusivity.

The strain rate is assumed to follow a Gaussian profile of fixed width $h$ across the fault \citep{garagash12}:
\begin{linenomath}
\begin{equation}\label{eq:gdot}
  \dot{\gamma} = \frac{V}{h}\exp\left(-\frac{\pi y^2}{h^2}\right),
\end{equation}
\end{linenomath}
where $V$ is the slip rate. The governing equation \eqref{eq:dpdt} requires a local evolution of porosity at spatial position $y$, and we assume here that the local porosity rate $\partial n^\mathrm{pl}/\partial t$ is directly proportional to the strain rate, with a proportionality factor given by the derivative of \eqref{eq:Dphi} with respect to slip:
\begin{linenomath}
\begin{equation}\label{eq:dnpldt}
  \frac{\partial n^\mathrm{pl}}{\partial t}  = \Delta\Phi_\mathrm{max}(h/\delta_\mathrm{D})\dot{\gamma}e^{-\delta/\delta_\mathrm{D}}.
\end{equation}
\end{linenomath}
The formulation \eqref{eq:dnpldt}, together with \eqref{eq:gdot}, is consistent with expression \eqref{eq:Dphi} for the spatially averaged porosity across the fault. Note that the product $\Delta\Phi_\mathrm{max}\times h$ gives the maximum total fault zone opening due to inelastic dilation.

The solution for pore pressure at $y=0$ as a function of time is given in integral form as \citep{garagash12}
\begin{linenomath}
\begin{equation}\label{eq:p(t)}
  p(t) - p_0 = \frac{\Lambda}{\rho c h}\int_0^t\tau(t')V(t')\mathcal{K}\left(\frac{t-t'}{T^*}; \frac{\alpha_\mathrm{hy}}{\alpha_\mathrm{th}}\right)dt' - \frac{1}{\beta^*}\int_0^tV(t')\Delta\Phi'[\delta(t')]\mathcal{A}\left(\frac{t-t'}{T_\mathrm{hy}}\right)dt',
\end{equation}
\end{linenomath}
where $\Delta\Phi'$ is the derivative \eqref{eq:Dphi} with respect to slip. The kernels $\mathcal{K}$ and $\mathcal{A}$ are given by \citet[][, Appendix A]{garagash12}. The coupling between pore pressure and temperature introduces two diffusion timescales, a thermo-hydraulic one, $T^*=h^2/4\alpha$, where $\alpha=(\sqrt{\alpha_\mathrm{hy}}+\sqrt{\alpha_\mathrm{th}})^2$, and a purely hydraulic one, $T_\mathrm{hy} = h^2/4\alpha_\mathrm{hy}$. In addition, thermal pressurisation introduces a natural weakening length scale which can be chosen as $\delta_\mathrm{c}=\rho c h/(f_\mathrm{r}\Lambda)$. Here, some flexibility exists in the definition of $\delta_\mathrm{c}$ since friction coefficient is not constant; the choice of $f_\mathrm{r}$ is made to model the case when thermal pressurisation becomes significant only at slip distances greater than $\delta_\mathrm{w}$, beyond which friction is in its weakened state \citep[a similar assumption was used by][in the case of early weakening by flash heating]{noda09,viesca15}.



\section{Dilatancy toughening of slip-weakening faults}


Let us first consider the rupture problem without thermal pressurisation, by assuming that $\Lambda$ is very small, so that the slip weakening distance $\delta_\mathrm{c}$ is much larger than both $\delta_\mathrm{w}$ and $\delta_\mathrm{D}$. Such a situation would arise in fault zones with large pore space compressibility, which seems to be the case of fresh granitic faults as reported by \citet{brantut20}.

\subsection{Effect of dilatancy on crack tip stress and slip rate}

The coupled problem \eqref{eq:tauf}, \eqref{eq:p(t)} and \eqref{eq:elastodyn} can be reformulated in terms of normalised quantities $x/L_\mathrm{w}$, $\Delta\tau/\tau_\mathrm{r}$, $V/V_\mathrm{w}$ and $\delta/\delta_\mathrm{w}$, where $\Delta\tau = \tau - \tau_\mathrm{r}$ is the stress drop and 
\begin{linenomath}
\begin{equation}
  \tau_\mathrm{r} = f_\mathrm{r}\sigma'_0, \quad L_\mathrm{w} = \frac{\bar{\mu}\delta_\mathrm{w}}{\tau_\mathrm{r}}, \quad V_\mathrm{w}=v_\mathrm{r}\delta_\mathrm{w}/L_\mathrm{w}.
\end{equation}
\end{linenomath}
The initial effective stress is denoted $\sigma'_0=\sigma_\mathrm{n}-p_0$. There are 4 parameters controlling the behaviour of the system, the ratios
\begin{linenomath}
\begin{equation}
  \frac{\delta_\mathrm{D}}{\delta_\mathrm{w}},\quad f_\mathrm{p}/f_\mathrm{r}, \quad \frac{\Delta\Phi_\mathrm{max}}{\beta^*\sigma'_0}, \quad\text{and } \frac{v_\mathrm{r}T_\mathrm{hy}}{L_\mathrm{w}} = \frac{h^2v_\mathrm{r}\tau_\mathrm{r}}{4\alpha_\mathrm{hy}\bar{\mu}\delta_\mathrm{w}} = \frac{T_\mathrm{hy}}{T},
\end{equation}
\end{linenomath}
where we denote $T = L_\mathrm{w}/v_\mathrm{r}$. The ratio $T_\mathrm{hy}/T = v_\mathrm{r}T_\mathrm{hy}/L_\mathrm{w}$ compares the hydraulic diffusion time to the characteristic time over which the slip-weakening region propagates along the fault, and is the key parameter controlling the transition from drained (small $T_\mathrm{hy}/T$) to undrained (large $T_\mathrm{hy}/T$) conditions near the rupture tip.

\begin{figure}
  \centering
  \includegraphics{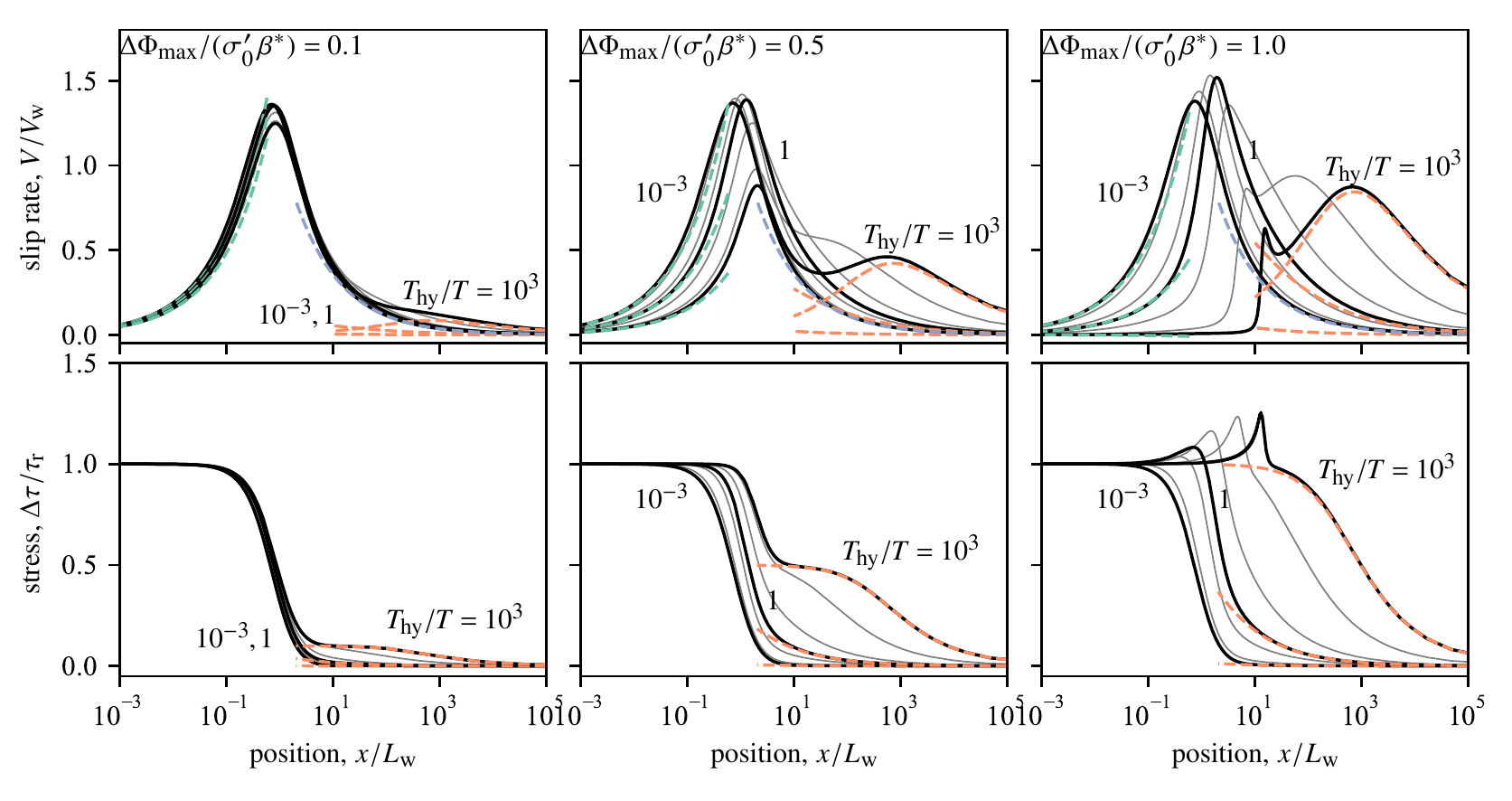}
  \caption{Slip rate (top) and strength (bottom) as a function of distance from the tip of a dynamically propagating crack for a range of undrained pore pressure change, parameter $\Delta\Phi_\mathrm{max}/(\beta^*\sigma_0')$, and hydraulic diffusion time, parameter $T_\mathrm{hy}/T$. Solutions are plotted for $\times 10$ increments in parameter $T_\mathrm{hy}/T$. Green dashed lines show the asymptotic behaviour $V(x)/V_\mathrm{w}= k\times (x/L_\mathrm{w})^{1/2}$, where $k$ is constant that depends on the undrained strength and is obtained numerically for each solution; Blue dashed line is the LEFM asymptote $V(x)/V_\mathrm{w}= 2/\sqrt{\pi}\times(x/L_\mathrm{w})^{-1/2}$ for the drained (dry) tip behaviour; Orange dashed lines correspond to the purely diffusive asymptote given in Equations \eqref{eq:tau_diff} and \eqref{eq:V_diff}.}
  \label{fig:swdil_soln}
\end{figure}

The solution is determined using the quadrature method detailed in \citet{viesca18}. A set of solutions is presented in Figure \ref{fig:swdil_soln}, where constant $\delta_\mathrm{D}/\delta_\mathrm{w}=1$ and $f_\mathrm{p}/f_\mathrm{r}=2$ were chosen as representative values. For significant values of $\Delta\Phi/\beta^*\sigma'_0$ (i.e., a nonnegligible undrained de-pressurisation due to dilatancy), we observe a clear transition from a drained regime at $T_\mathrm{hy}/T\ll 1$, where the crack tip behaviour is essentially governed by the dry slip weakening mechanisms, to a dual scale behaviour at $T_\mathrm{hy}/T\gg 1$ where the tip is undrained, governed by a slip-dependent behaviour (combination of slip weakening and undrained effective stress law), and an extended region (at $x>v_\mathrm{r}T_\mathrm{hy}$) governed by the progressive pore pressure recharge of the fault zone from the off fault regions (see also Figure \ref{fig:setup}b). This second weakening behaviour is purely diffusive, and at $x\gg v_\mathrm{r}T_\mathrm{hy}$ the strength is well approximated by (this follows from approximation of the second integral in \eqref{eq:p(t)}, assuming concentrated source at $t=0$):
\begin{linenomath}
\begin{equation} \label{eq:tau_diff}
\Delta\tau(x)/\tau_\mathrm{r} \approx \frac{\Delta\Phi_\mathrm{max}/(\beta^*\sigma'_0)}{\sqrt{1 + \pi x/(v_\mathrm{r}T_\mathrm{hy})}}, 
\end{equation}
\end{linenomath}
and the slip rate behaves as (this follows from inversion of the one-sided Hilbert transform in \eqref{eq:elastodyn} and direct computation of the resulting integral)
\begin{linenomath}
\begin{equation} \label{eq:V_diff}
 V(x)/V_\mathrm{w} \approx \frac{4}{\pi}\frac{\Delta\tau(x)}{\tau_\mathrm{r}}\mathrm{asinh}\left(\sqrt{\pi x/(v_\mathrm{r}T_\mathrm{hy})}\right).
\end{equation}
\end{linenomath}
With increasing undrained pore pressure change (parameter $\Delta\Phi_\mathrm{max}/(\beta^* \sigma'_0)$), the peak slip rate at the crack tip decreases, due to a smaller net strength drop under undrained conditions, while the slip rate in the extended ``far-tip'' region controlled by diffusive pore pressure recharge increases.

\subsection{Stress intensity factor and fracture energy: Small scale yielding approximation}

\begin{figure}
  \centering
  \includegraphics{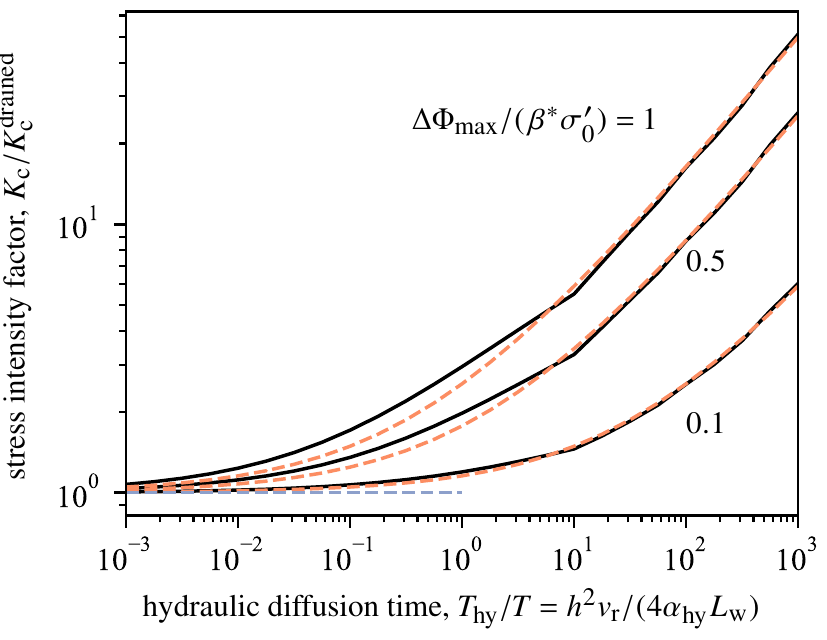}
  \caption{Stress intensity factor required to drive crack expansion, normalised by its perfectly drained value, as a function of hydraulic diffusion time relative to characteristic frictional weakening time. The cutoff length was set to $\ell_\mathrm{cut} = \mathrm{max}\{100L_\mathrm{w}, 10\times v_\mathrm{r}T_\mathrm{hy}\}$ in the numerical computation. $T_\mathrm{hy}/T$ increases with increasing rupture speed. Blue dashed line indicates the purely drained limit. Orange dashed lines show the analytical estimate as computed from the sum of the drained end-member and the large $T_\mathrm{hy}/T$ end-member \eqref{eq:K_diff}, using $\ell_\mathrm{cut}=10\times v_\mathrm{r}T_\mathrm{hy}$.}
  \label{fig:Kc}
\end{figure}

Now equipped with a numerical solution and relevant asymptotes for the stress distribution along the moving crack, we can turn our attention to the desired analysis in terms of integrated fracture mechanics quantities. By construction, there is no stress or slip rate singularity at the tip of our moving crack. However, we can consider the situation where all the strength change due to the coupled slip weakening and dilation process occurs over a length scale much smaller than the finite size of the expanding crack. In that case, the stress distribution corresponds to a stress intensity factor that needs to be overcome for the crack to keep expanding, which is given by \citep{kostrov66}
\begin{linenomath}
\begin{equation} \label{eq:SIFkostrov}
  K_\mathrm{c} = \sqrt{1-v_\mathrm{r}/c_\mathrm{s}}\sqrt{\frac{2}{\pi}}\int_0^{c_\mathrm{s}t} \frac{\Delta\tau(x, t-x/c_\mathrm{s})}{\sqrt{x}}dx.
\end{equation}
\end{linenomath}
In the  quasi-static case ($v_\mathrm{r}\ll c_\mathrm{s}$), it simplifies to \citep[][pp. 598, 604]{rice80}
\begin{linenomath}
\begin{equation}
  \label{eq:SIF1}
  K_\mathrm{c} = \sqrt{\frac{2}{\pi}}\int_0^\infty \frac{\Delta\tau(x)}{\sqrt{x}}dx. 
\end{equation}
\end{linenomath}
The purely drained end-member case, where $\Delta\tau$ is an exponentially decreasing function of slip, provides a lower bound $K_\mathrm{c}^\mathrm{drained}$ for the stress intensity factor. This lower bound is obtained from the finite fracture energy associated with the exponential slip-weakening process ($G_\mathrm{c}^\mathrm{drained}=\delta_\mathrm{w}\sigma'_0(f_\mathrm{p}-f_\mathrm{r})$):
\begin{linenomath}
\begin{equation}\label{eq:Kcdry}
  K_\mathrm{c}^\mathrm{drained} = \sqrt{2\bar{\mu}G_\mathrm{c}^\mathrm{drained}} = \tau_\mathrm{r}\sqrt{2L_\mathrm{w}(f_\mathrm{p}/f_\mathrm{r}-1)},
\end{equation}
\end{linenomath}
where the relationship between stress intensity factor $K$ and energy release rate $G$
\begin{linenomath}
\begin{equation}
  K^2/2\bar{\mu} = G
\end{equation}
\end{linenomath}
was used. When dilatancy produces a significant pore pressure drop, the stress in the vicinity of the rupture tip increases, and thus $K_\mathrm{c}$ increases. The diffusive nature of the stress evolution introduces some surprising complexity in the problem: the integral of \eqref{eq:SIF1} does not converge. Following \citet{rice73}, a cutoff length $\ell_\mathrm{cut}$ is introduced beyond which the strength is considered to be exactly zero, in accordance with the original assumption that all the strength evolution remains confined to a small region near the crack tip. The stress intensity factor computed this way exhibits a marked increase with increasing diffusion time relative to frictional weakening time $T_\mathrm{hy}/T$, that is, with increasing rupture speed (Figure \ref{fig:Kc}). Therefore, dilatancy produces an effective toughening of the fault, which potentially limits the rupture speeds. Using the approximation \eqref{eq:tau_diff} for the stress distribution near the crack tip, the dynamic stress intensity factor is approximated by
\begin{linenomath}
\begin{equation} \label{eq:K_diff}
  K_\mathrm{c} \approx 2\sqrt{\frac{2}{\pi}}\tau_\mathrm{r}\frac{\Delta\Phi_\mathrm{max}}{\beta^*\sigma'_0}\sqrt{v_\mathrm{r}T_\mathrm{hy}/\pi} \mathrm{asinh}\left(\sqrt{\pi \ell_\mathrm{cut}/(v_\mathrm{r}T_\mathrm{hy})}\right)\qquad\text{(large }T_\mathrm{hy}/T\text{)}.
\end{equation}
\end{linenomath}
This expression for stress intensity factor has a similar behaviour to that derived by \citet{rice73} for the case of an infinitesimally thin slipping zone and quasi-static rupture growth, where $K$ was found to increase as $\sqrt{v_\mathrm{r}}$, and a weak dependence on $\ell_\mathrm{cut}$ was established. 

\begin{figure}
  \centering
  \includegraphics{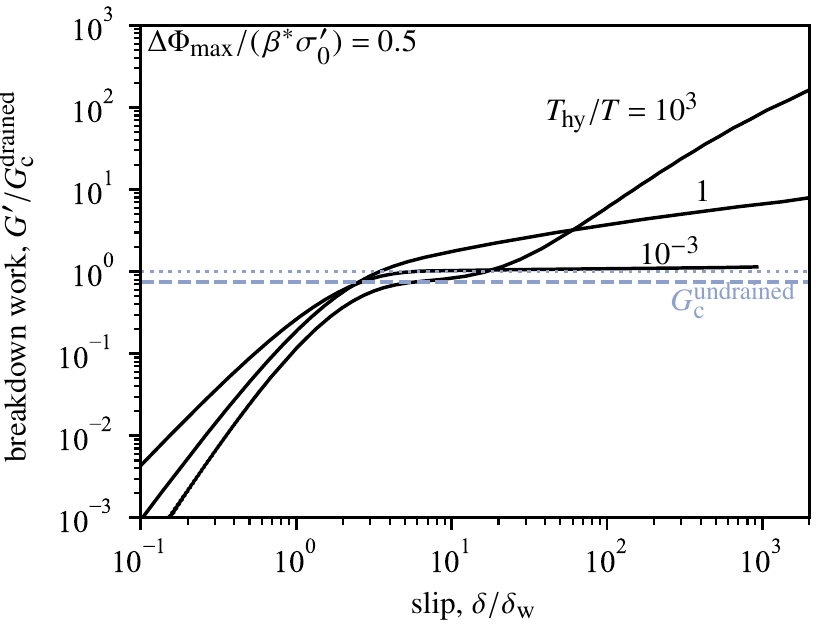}
  \caption{Breakdown work, normalised by its finite drained value, as a function of slip, for dynamic semi-infinite crack models computed using $f_\mathrm{p}/f_\mathrm{r}=2$, $\Delta\Phi_\mathrm{max}/(\beta^*\sigma'_0)=0.5$ and $\delta_\mathrm{D}=\delta_\mathrm{w}$.}
  \label{fig:G}
\end{figure}

A complementary approach to computing a critical stress intensity factor is to estimate the equivalent fracture energy associated with the weakening processes at the crack tip. Following \citet{rice06}, we can use a generalised form of fracture energy, better called ``breakdown work'', defined as
\begin{linenomath}
\begin{equation}\label{eq:G'}
  G'(\delta) = \int_0^\delta\tau(\delta')-\tau(\delta)d\delta',
\end{equation}
\end{linenomath}
which reduces to the conventional form established by \citet{palmer73} when a well defined constant residual stress exists and slip is large compared to any characteristic slip-weakening distance. Under drained conditions, for instance at very small rupture speed ($T_\mathrm{hy}/T\ll1$), $G'$ becomes almost constant at slip distances that are large compared to $\delta_\mathrm{w}$ (Figure \ref{fig:G}). In that case, $G'\approx G_\mathrm{c}^\mathrm{drained}$ and there is a well defined cohesive zone of small dimensions. With decreasing drainage (increasing $T_\mathrm{hy}/T$), $G'$ evolves in two distinct regimes: (1) an increase and stabilisation up to slip distances of the order of $\delta_\mathrm{w}$, and (2) a marked increase with increasing slip beyond $\delta_\mathrm{w}$. The first regime corresponds to a purely undrained behaviour, where shear strength evolves as
\begin{linenomath}
\begin{equation}
  \tau_\mathrm{f} = \left(f_\mathrm{r}+(f_\mathrm{p}-f_\mathrm{r})e^{-\delta/\delta_\mathrm{w}}\right)\left(\sigma'_0 + (\Delta\Phi_\mathrm{max}/\beta^*)(1-e^{-\delta/\delta_\mathrm{D}})\right)\qquad\text{(undrained)}.
\end{equation}
\end{linenomath}
The undrained behaviour at the tip is associated to a well defined, constant fracture energy
\begin{linenomath}
\begin{equation}\label{eq:Gcundrained}
  G_\mathrm{c}^\mathrm{undrained} = (f_\mathrm{p}/f_\mathrm{r}-1)\tau_\mathrm{r}\delta_\mathrm{w}\left(1+\frac{\Delta\Phi_\mathrm{max}/(\beta^*\sigma'_0)}{\delta_\mathrm{D}/\delta_\mathrm{w}+1}\right) - f_\mathrm{r}\delta_\mathrm{D}\Delta\Phi_\mathrm{max}/\beta^*,
\end{equation}
\end{linenomath}
and the subsequent increase occurs due to diffusive recharge of the fault zone. The breakdown work, as computed by \eqref{eq:G'}, does not converge to a fixed value at large slip: this is consistent with the theoretically infinite toughness and the requirement of a cutoff length $\ell_\mathrm{cut}$ that limits the size of the cohesive zone. The unlimited increase in $G'$ exhibited by our slip-weakening, dilatant fault zone model, is a general feature of weakening models limited by hydro-thermal diffusion processes \citep{rice06,viesca15,brantut17b}.

For rupture propagating in consolidated rocks, using representative parameter values of $\delta_\mathrm{w}\approx\delta_\mathrm{D}\approx 1$~mm, $\mu \approx 30$~GPa, $(f_\mathrm{p}/f_\mathrm{r}-1)\tau_\mathrm{r}\approx 30$~MPa and $\alpha\approx10^{-5}$~m$^2$s$^{-1}$, the key controlling parameter $T_\mathrm{hy}/T$ ranges from around $10^{-4}$ at $v_\mathrm{r}\sim1$~km/day up to $10$ at $v_\mathrm{r}\sim2$~km/s when considering a relatively thin dilatant zone of $h=1$~mm, and from $1$ to $10^5$ over the same rupture speed range when considering a thicker zone of $h=10$~cm. For $v_\mathrm{r}$ approaching $c_\mathrm{s}$, $\bar{\mu}$ vanishes and thus $T_\mathrm{hy}/T$ becomes unbounded. Therefore, for thin shear zones, dilatancy is expected to produce dramatic toughening only at dynamic rupture speeds, whereas large toughening is predicted even at small rupture speeds, of the order of 10 km/day, if dilatancy occurs in thick zones.

\subsection{Stress intensity factor and fracture energy: ``near'' and ``far''-tip contributions}
\label{sec:energy}

While the toughening effect is somewhat intuitive, one caveat is that the concept of toughness and the relevance of the stress intensity factor to characterise crack expansion may be limited, since it relies on the small scale yielding assumption: in computing $K_\mathrm{c}$ and using it in terms of ``crack resistance'', one has to assume that the nonlinear end region, where the stress evolves from the nominal strength to the residual value, is small compared to the overall crack size. The need for a cutoff length $\ell_\mathrm{cut}$ to prevent the mathematical divergence of the integral in \eqref{eq:SIF1} indicates that the nonlinear end region may not be small in general: it is only approximately true when the rupture dimensions are much larger than the drainage length scale ($v_\mathrm{r}T_\mathrm{hy}$).

In practice, for finite ruptures, the integral would extend along the full length of the rupture, making the toughness size-dependent (with longer ruptures having larger toughness than short ones). Size-dependent toughness is consistent with the results depicted in Figure \ref{fig:G} showing slip-dependent breakdown work: longer ruptures, which typically correspond to larger slip, have larger breakdown work. Such a size-dependence of toughness is commonly observed in engineering materials due to so-called crack bridging effects \citep[e.g.,][]{cox94}. Here the matter is further complicated by the time-dependence of the strength evolution, which would make the toughness time-dependent as well. 

An alternative to grouping all nonlinearities in the toughness and making it size- and time-dependent, is to separate the various contributions to stress intensity factor occurring at different scales. If some of the strength drop is not confined to a small region near the crack tip, which is the case when the rupture dimension is comparable to or smaller than the diffusion length scale, it will not contribute to toughness (or fracture energy) \textit{per se}, but to the stress intensity factor (or energy release rate). Here, it is natural to have the nonlocal contribution of diffusive fault zone recharge into $G$, and keep the undrained contribution of dilatancy into $G_\mathrm{c}$. In any case, the energy balance $G=G_\mathrm{c}$ is always verified, but $G$ now includes nonlinearities due to the constitutive behaviour of the fault.

A simple illustrative example can be worked out: consider a semi-infinite crack along direction $x$, propagating at constant rupture speed $v_\mathrm{r}$ in a background stress field given by $\tau_\mathrm{b}>0$ for $x\geq0$ and $\tau_\mathrm{b}=0$ for $X<0$. The stress drop along the crack is $\tau_\mathrm{b}-\tau_\mathrm{f}$, and for sufficiently large diffusion time, the strength evolution $\tau_\mathrm{f}$ can be split into a near tip, undrained contribution, and a spatially extended diffusive contribution. The near tip undrained contribution corresponds to an undrained toughness $K_\mathrm{c}^\mathrm{undrained} = \sqrt{2\bar{\mu}G_\mathrm{c}^\mathrm{undrained}}$. This toughness must be matched by the stress intensity factor associated with the background loading, offset by the spatially extended diffusive contribution \eqref{eq:tau_diff}. This stress intensity factor, computed from \eqref{eq:SIFkostrov}, yields
\begin{linenomath}
\begin{equation} \label{eq:Kdyn}
  K/\sqrt{1-v_\mathrm{r}/c_\mathrm{s}} = \sqrt{\frac{2}{\pi}}\left[2\sqrt{\ell}(\tau_\mathrm{b}-\tau_\mathrm{r}) - 2 (f_\mathrm{r}\Delta\Phi_\mathrm{max}/\beta^*)\sqrt{v_\mathrm{r}T_\mathrm{hy}/\pi}\mathrm{asinh}\left(\pi\ell/(v_\mathrm{r}T_\mathrm{hy})\right)\right].
\end{equation}
\end{linenomath}
In the limit $\ell\ll v_\mathrm{r}T_\mathrm{hy}$, the stress intensity factor is simply:
\begin{linenomath}
\begin{equation} \label{eq:K_early}
  K/\sqrt{1-v_\mathrm{r}/c_\mathrm{s}} \approx \frac{2}{\sqrt{\pi}}\sqrt{2\ell}\left[\tau_\mathrm{b} - f_\mathrm{r}\Delta\Phi_\mathrm{max}/\beta^* - \tau_\mathrm{r}\right], \quad\text{(}\ell\ll v_\mathrm{r}T_\mathrm{hy}\text{)}
\end{equation}
\end{linenomath}
so that dilatancy has the effect of reducing the stress drop, and limiting the available energy for crack tip extension. By contrast, at large rupture dimensions ($\ell\gg v_\mathrm{r}T_\mathrm{hy}$), the stress intensity factor becomes
\begin{linenomath}
\begin{equation}
  K/\sqrt{1-v_\mathrm{r}/c_\mathrm{s}} \approx \frac{2}{\sqrt{\pi}}\sqrt{2\ell}\left(\tau_\mathrm{b} - \tau_\mathrm{r}\right) - \left(f_\mathrm{r}\Delta\Phi_\mathrm{max}/\beta^* \right)\sqrt{2v_\mathrm{r}T_\mathrm{hy}}\ln\left(2\sqrt{\ell/(v_\mathrm{r}T_\mathrm{hy})}\right), \quad\text{(}\ell\gg v_\mathrm{r}T_\mathrm{hy}\text{)}
\end{equation}
\end{linenomath}
and the contribution of dilatancy eventually becomes negligible compared to that of the applied load.

The simple example of a semi-infinite crack propagating at constant speed is clearly not a fully realistic earthquake or slow-slip model, and the choice of applied load was made for mere mathematical convenience. However, it is quite illuminating to establish that dilatancy has a two-fold effect: (1) changing the fracture energy at the crack tip, and (2) reducing the stress drop and therefore the energy release rate in a time-dependent, diffusive manner. The different contributions of dilatancy and drainage can be recast in the Griffith energy balance,
\begin{linenomath}
\begin{equation} \label{eq:griffith}
  G = K^2/2\bar{\mu} = G_\mathrm{c}^\mathrm{undrained},
\end{equation}
\end{linenomath}
which provides an equation of motion for the crack tip through the dependency of $G$ on $v_\mathrm{r}$ and crack tip position. The energy release rate $G$ is thus impacted by the combination of applied load and diffusive pore pressure recharge along the crack, while the undrained behaviour (combination of slip weakening and pore pressure drop) controls fracture energy. The energy release rate depends quadratically on $K$, and is therefore not simply offset by the diffusive pore pressure recharge far from the tip. Thus, in general, the full contribution of dilatancy and drainage to the dynamics of rupture is not simply a change in fracture energy. The approximation of small scale yielding made in the previous section (see Equation \ref{eq:K_diff}), which followed the developments of \citet{rice73}, is thus valid only in the limit of ruptures that extend well beyond the region impacted by pore pressure recharge.

\begin{figure}
  \centering
  \includegraphics{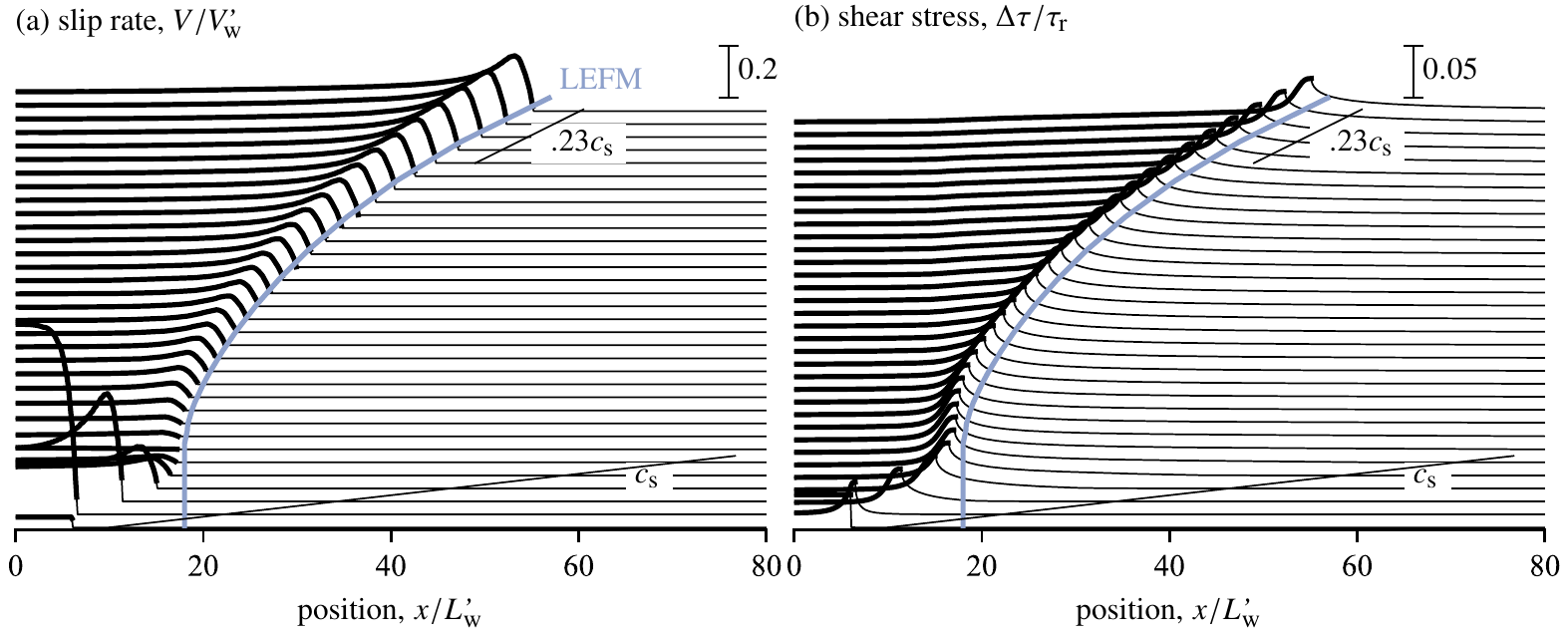}
  \caption{Slip rate (a) and stress (b) profiles during an elastodynamic rupture simulation with $\Delta\Phi_\mathrm{max}/(\sigma'_0\beta^*)=0.5$, $\delta_\mathrm{w}=\delta_\mathrm{D}$, $f_\mathrm{p}=f_\mathrm{r}$, and a normalised hydraulic diffusion time $T_\mathrm{hy}/T'=10^3$, where the characteristic frictional weakening time is defined as $T' = L_\mathrm{w}'/c_\mathrm{s}$, the characteristic distance is $L_\mathrm{w}'=\delta_\mathrm{w}\tau_\mathrm{r}/\mu$, and the characteristic slip rate is $V_\mathrm{w}'=\delta_\mathrm{w}/T'$. The background stress was uniform at $1.5\tau_\mathrm{r}$, except for a small region of size $12L_\mathrm{w}'$ near the origin where its value was $1$\% above the peak strength. The rupture is symmetric with respect to $x=0$ and only the domain $x>0$ is displayed. Profiles are plotted at time intervals equal to $12.5T'$. Blue lines in each plot show the prediction of rupture tip trajectory from the Griffith energy balance.}
  \label{fig:dilexample1}
\end{figure}

\subsection{Comparison to dynamic simulations}

Whether due to a direct toughening effect, or to a reduction in energy release rate, dilatancy is expected to slow down rupture propagation. This can be tested in full elastodynamic simulations (see details in Appendix \ref{ax:dyn}). An illustrative example is shown in Figure \ref{fig:dilexample1}, where key constitutive parameters were set to $\Delta\Phi_\mathrm{max}/(\sigma'_0\beta^*)=0.5$, $f_\mathrm{p}/f_\mathrm{r}=2$, and the background stress outside the nucleation zone was set to $0.5\tau_\mathrm{r}$ above the residual frictional strength. In that case, a purely undrained behaviour would amount to zero stress drop, so that rupture should not grow unless drainage becomes significant ($G$, or $K$, is exactly zero under purely undrained conditions, cf. Equation \ref{eq:K_early}). In the simulation, the rupture tip is indeed arrested at early times, at around $x/L_\mathrm{w}'=18$ (where $L_\mathrm{w}'=\mu\delta_\mathrm{w}/\tau_\mathrm{r}$). The only mechanism that allows rupture to grow any further is the diffusive pore pressure recharge of the fault zone, which increases the strength drop in the interior of the crack, and increases the energy release rate. After its momentary arrest, rupture slowly accelerates, but remains well below the shear wave speed (the limiting speed in mode III) over the full simulation time, of the order of $400\times \mu\delta_\mathrm{w}/(\tau_\mathrm{r}c_\mathrm{s})$. By comparison, simulated ruptures in the same background stress but without dilatancy accelerate to $v_\mathrm{r}\approx c_\mathrm{s}$ almost immediately after nucleation.

\begin{figure}
  \centering
  \includegraphics{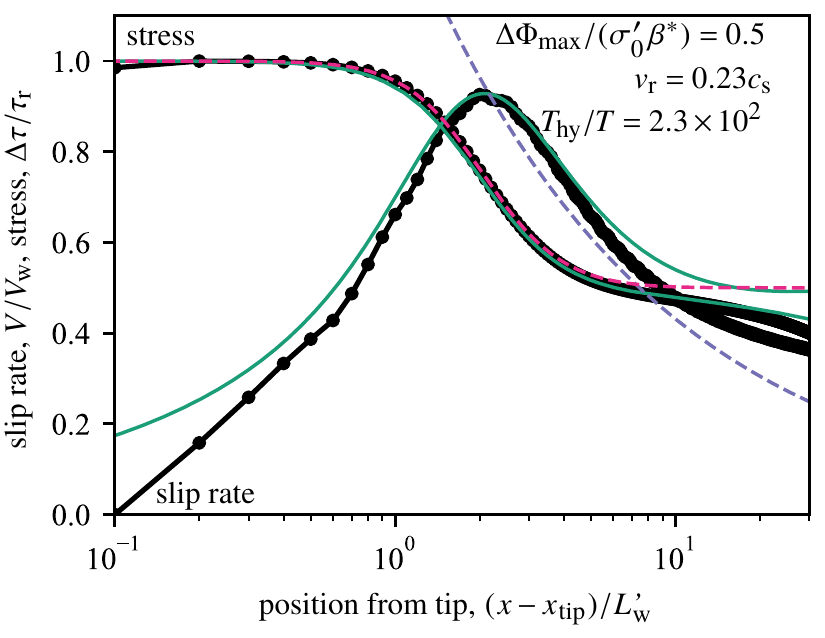}
  \caption{Slip rate and stress profile behind the crack tip during the last stage of the dynamic rupture shown in Figure \ref{fig:dilexample1}, where rupture speed is approximately equal to $0.23c_\mathrm{s}$ (black line and points), and steady-state semi-infinite crack solution (green) for the same parameter values. The blue dashed line is the LEFM prediction $V/V_\mathrm{w} = 2K_\mathrm{c}^\mathrm{undrained}/\sqrt{\pi}\times(x/L_\mathrm{w})^{-1/2}$, and the red dashed line is the undrained strength prediction.}
  \label{fig:comparison1}
\end{figure}

The stress and slip rate near the tip of the spontaneous dynamic rupture are well characterised by the steady-state solution  using the appropriate transient rupture speed (Figure \ref{fig:comparison1}). The tip is indeed under essentially undrained conditions, and the slip rate is reasonably well characterised by the linear elastic fracture mechanics (LEFM) solution $V/V_\mathrm{w}= 2K_\mathrm{c}^\mathrm{undrained}/\sqrt{\pi}\times(x/L_\mathrm{w})^{-1/2}$, corresponding to the undrained stress intensity factor as defined by $K_\mathrm{c}^\mathrm{undrained}=\sqrt{2\bar{\mu}G_\mathrm{c}^\mathrm{undrained}}$. In the dynamic model, the progressive strength drop associated with drainage does not lead to a significant increase in slip rate further away from the tip (as predicted by the semi-infinite crack solution) due to the finite crack size of the model. 



The usefulness of the LEFM approximation can be tested by comparing the rupture tip trajectory obtained from the full numerical solution to that based on the crack tip equation of motion arising from the Griffith energy balance. In view of the undrained behaviour at the tip, it is appropriate to equate the dynamic energy release rate to the undrained fracture energy $G_\mathrm{c}^\mathrm{undrained}$, and to include the strength change due to pore fluid diffusion (given asymptotically by Equation \eqref{eq:tau_diff} far from the tip) in the stress drop that enters in the expression of $G$ (Equation \ref{eq:griffith}). The details of the solution are given in Appendix \ref{ax:eofm}. The rupture tip trajectory resulting from this procedure is in remarkable agreement with the full numerical solution (Figure \ref{fig:dilexample1}), illustrating the practical efficiency of the LEFM approximation.





\section{Link with thermal pressurisation}


In the previous Section we have established how dilatancy and pore fluid diffusion influence the dynamics of rupture, without regards to thermal effects. In general, thermal pressurisation is expected to produce significant weakening for slip distances larger than $\delta_\mathrm{c}$. In granite gouge, $\delta_\mathrm{c}$ is typically of the order of $10$~mm \citep{brantut16b}, which is around one order of magnitude larger than $\delta_\mathrm{w}$ or $\delta_\mathrm{D}$. In freshly fractured granite, \citet{brantut20} reported very high pore space compressibility compared to previously reported ``mature'' fault gouge material, which indicates that $\delta_\mathrm{c}$ is likely much larger than the estimate of $10$~mm in consolidated rocks.

In any case, thermal pressurisation is not just superimposed onto the slip weakening and dilatancy processes, but is coupled with them: as noticed by \citet{garagash03a}, the de-pressurisation effect of dilatancy could be compensated by increased shear heating and fluid pressurisation.

When dealing with the fully coupled slip weakening, dilatancy and thermal pressurisation dynamic problem, a total of 6 nondimensional quantities control the system's behaviour:
\begin{linenomath}
\begin{equation}
  \frac{\delta_\mathrm{D}}{\delta_\mathrm{w}},\quad f_\mathrm{p}/f_\mathrm{r}, \quad \frac{\Delta\Phi_\mathrm{max}}{\beta^*\sigma'_0}, \quad \frac{T_\mathrm{hy}}{T},\quad \chi=\frac{\alpha_\mathrm{hy}}{\alpha_\mathrm{th}},\,\text{and}\quad\frac{\delta_\mathrm{c}}{\delta_\mathrm{w}}=\frac{\rho c h}{f_\mathrm{r}\Lambda \delta_\mathrm{w}}.
\end{equation}
\end{linenomath}

Let us consider a representative case by choosing $\delta_\mathrm{D}/\delta_\mathrm{w}=1$, $f_\mathrm{p}/f_\mathrm{r}=1$, $\chi=10$, $\delta_\mathrm{c}/\delta_\mathrm{w}=10$, and examine how dilatancy and hydro-thermal diffusion influence rupture dynamics, solving the steady-state problem (Equation \eqref{eq:elastodyn} with \eqref{eq:p(t)}, \eqref{eq:f} and \eqref{eq:Dphi}). When thermal pressurisation is active, the residual strength is zero, so that the correct assumption to solve the semi-infinite rupture problem is that the applied stress is also zero.

\begin{figure}
  \centering
  \includegraphics{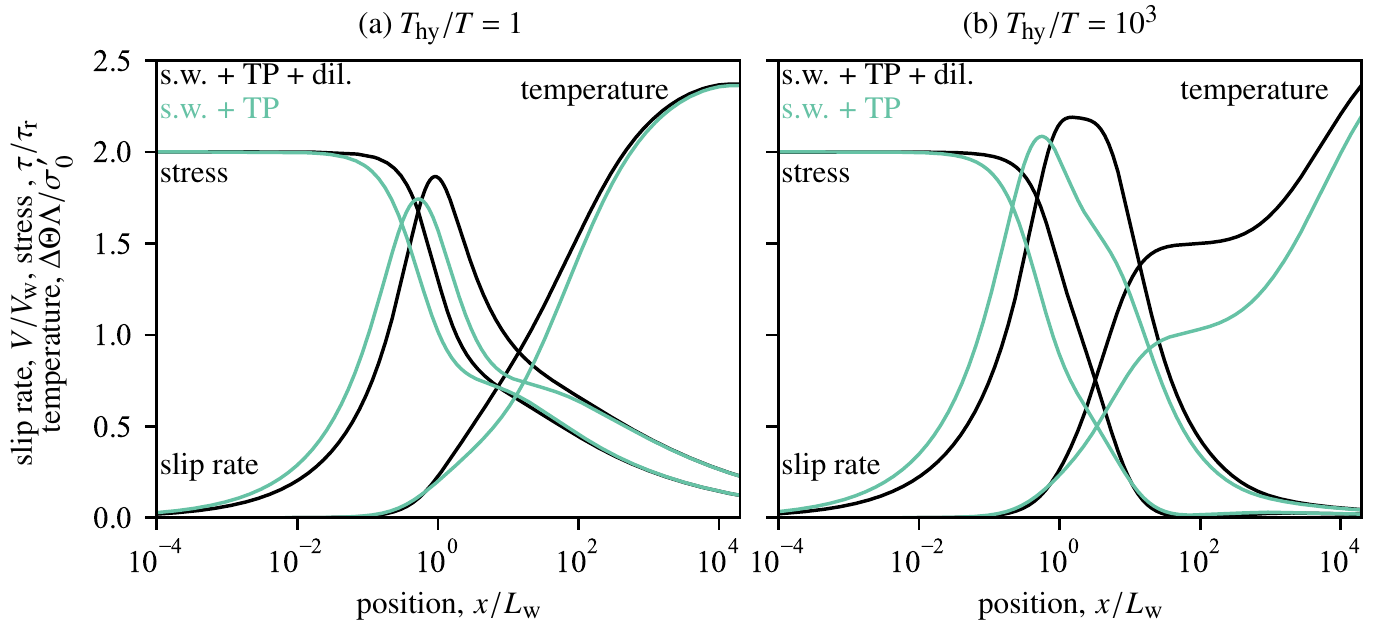}
  \caption{Slip rate, strength and temperature as a function of distance from the tip of a dynamically propagating crack driven by slip weakening friction, thermal pressurisation, with (black lines, $\Delta\Phi_\mathrm{max}/(\beta^*\sigma_0')=0.5$) and without (green lines, $\Delta\Phi_\mathrm{max}/(\beta^*\sigma_0')=0$) dilatancy. Hydraulic diffusion time is set to $T_\mathrm{hy}/T=1$ (a) and $T_\mathrm{hy}/T=10^3$ (b).}
  \label{fig:tpdil_soln}
\end{figure}

In the absence of dilatancy and with relatively fast diffusion ($T_\mathrm{hy}/T=1$; Figure \ref{fig:tpdil_soln}(a)), the stress evolution follows two stages of weakening, one associated with frictional weakening down to around $\tau/\tau_\mathrm{r}=1$ over a distance of the order of $L_\mathrm{w}$, and a subsequent, slow decrease down to near $0$ strength, associated with thermal pressurisation. This is the situation presented in detail by \citet[][their figure 1b]{viesca15}. When dilatancy is included, here choosing $\Delta\Phi_\mathrm{max}/(\beta^*\sigma'_0)=0.5$, a significant strengthening is observed and the first stage of weakening is effectively delayed. However, at large distance from the tip, strength, slip rate and temperature become indistinguishable from the nondilatant case. The weakening due to thermal pressurisation is indeed similar between the two cases due to (1) the extra heating rate (temperature is initially higher in the dilatant case) in the undrained regime, and (2) the diffusive pore pressure recharge from the off-fault region.

The same general pattern is observed when diffusion time is larger ($T_\mathrm{hy}/T=10^3$, Figure \ref{fig:tpdil_soln}(b)): an initial delayed weakening due to frictional slip weakening coupled to dilatancy, followed by weakening driven by thermal pressurisation with an increased heating rate compared to the nondilatant case. The effect of dilatancy vanishes at distances $x\gg v_\mathrm{r}T_\mathrm{hy}$. When $\delta_\mathrm{D}\ll\delta_\mathrm{c}$ (as in the case shown in Figure \ref{fig:tpdil_soln}), the adiabatic, undrained behaviour including dilatancy is well described by a simple resetting of the initial pore pressure by the quantity $\Delta\Phi_\mathrm{max}/\beta^*$ \citep[][\S 43]{rice06}; this is illustrated by the temperature rise, which is offset by $\Delta\Phi_\mathrm{max}/(\beta^*\Lambda)$ compared to the nondilatant case (see Appendix \ref{ax:TP_au}), in the domain $L_\mathrm{w}\delta_\mathrm{c}/\delta_\mathrm{w}<x<v_\mathrm{r}T_\mathrm{hy}$.


\begin{figure}
  \centering
  \includegraphics{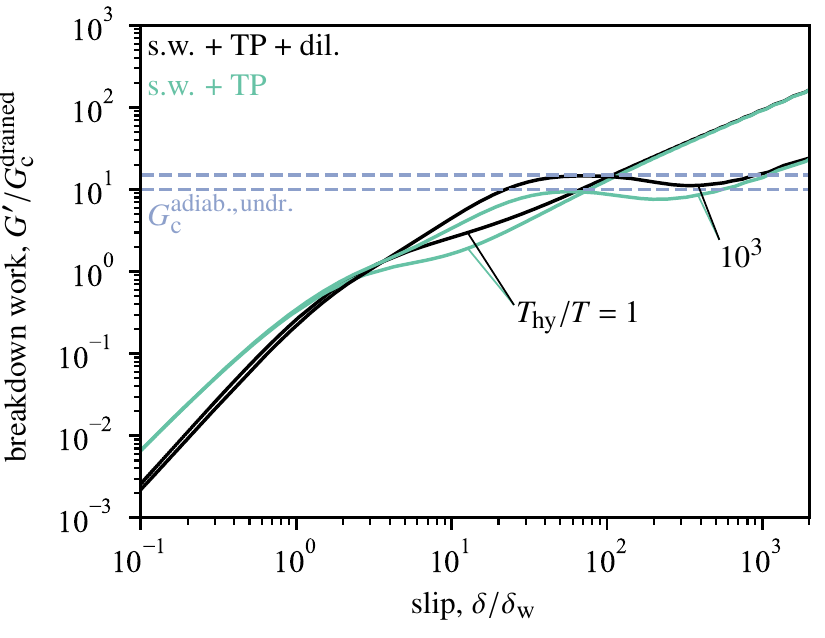}
  \caption{Breakdown work, normalised by its finite drained value, as a function of slip, for dynamic semi-infinite crack models computed using $f_\mathrm{p}/f_\mathrm{r}=2$, $\delta_\mathrm{D}/\delta_\mathrm{w}=1$, $\delta_\mathrm{c}/\delta_\mathrm{w}=10$, $\chi=10$, $T_\mathrm{hy}/T=1$ or $10^3$, and with (black lines, $\Delta\Phi_\mathrm{max}/(\beta^*\sigma_0')=0.5$) or without (green lines,  $\Delta\Phi_\mathrm{max}/(\beta^*\sigma_0')=0$) dilatancy. The dashed lines correspond to the fracture energy computed using \eqref{eq:GcTP} with and without dilatancy.}
  \label{fig:GTP}
\end{figure}

The main effect of dilatancy when coupled to thermal pressurisation is a relative strengthening near the crack tip. This near-tip strengthening corresponds to an increase in breakdown work at small slip distances (Figure \ref{fig:GTP}). Under purely undrained, adiabatic conditions (near the crack tip), there is a well defined fracture energy (the integral \eqref{eq:G'} converges). For constant friction coefficient, the fracture energy is simply \begin{linenomath}
\begin{equation}\label{eq:GcTP}
  G_\mathrm{c}^\mathrm{adiab., undr.} = \tau_\mathrm{r}\delta_\mathrm{c}\left(1+\Delta\Phi_\mathrm{max}/(\beta^*\sigma_0')\right),\quad\text{(constant friction)},
\end{equation}
\end{linenomath}
where it is easily seen how dilatancy is merely offsetting the effective stress. The fracture energy \eqref{eq:GcTP} is a good approximation for the breakdown work at small slip when diffusivity is low (Figure \ref{fig:GTP}, dashed lines). For large rupture dimensions and large slip distances, hydro-thermal diffusion becomes significant and the effect of dilatancy on strength and breakdown work becomes progressively negligible (see curves merging in Figures \ref{fig:tpdil_soln} and \ref{fig:GTP}).

Similarly to the situation where thermal pressurisation was neglected, the continuous increase in breakdown work with increasing slip, at large distances from the crack tip, indicates that the weakening may not be confined to a well-defined small region near the crack tip. Thus, the breakdown work should not be given the meaning of ``fracture energy'' in the sense that it may not be used directly in a crack tip equation of motion of the kind $G=G_\mathrm{c}$. Rather, as discussed in the previous section, if the weakening scales associated with friction, dilatancy and undrained, adiabatic thermal pressurisation are much smaller than the length scale associated with hydro-thermal diffusion, one may separate the tip behaviour and use the small scale yielding approximation to determine a constant fracture energy (e.g., $G_\mathrm{c}^\mathrm{adiab., undr.}$), and the continued weakening far from the crack tip will then contribute to the stress intensity factor (or energy release rate).

Because dilatancy is expected to occur primarily over the first few millimetres of slip, concomitant with any frictional weakening process, its contribution is significant in the near-tip region. The key effect of dilatancy  is therefore to increase the undrained fracture energy, which has the potential to locally slow down or stop rupture propagation.

\section{Discussion}


\subsection{Onset of vaporisation and melting}

Throughout this paper the properties of the pore fluid have been assumed constant. This is a reasonable hypothesis when the fluid is a liquid, but the strong changes in pressure (and temperature) associated with dilatancy and thermal pressurisation can significantly impact the fluid compressibility, thermal expansivity and heat capacity. Such changes have been invoked to explain slip dynamics in water saturated granite \citep{acosta18}, and recent experimental data have brought direct evidence for fluid vaporisation (with dramatic increase in fluid compressibility) due to dilatancy during rock failure \citep{brantut20,aben21}.

\begin{figure}
  \centering
  \includegraphics{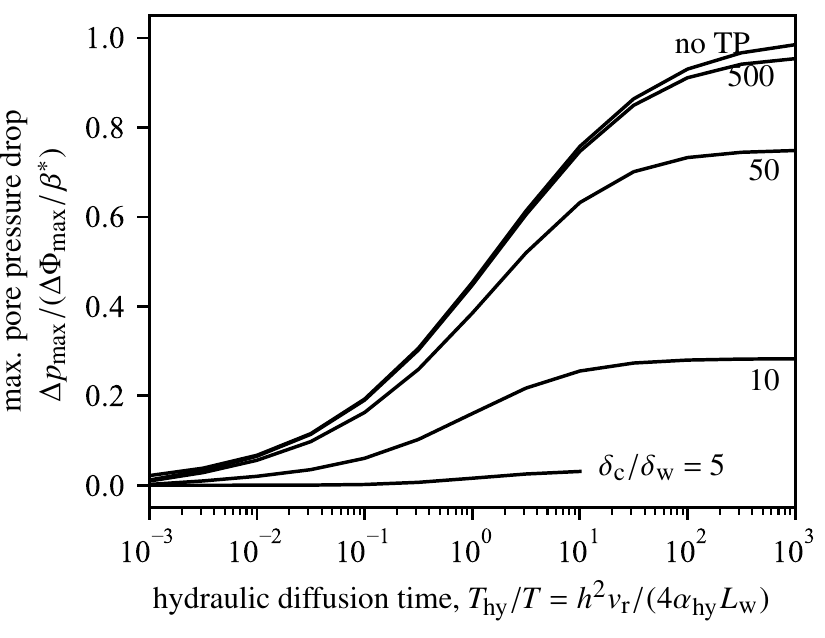}
  \caption{Maximum pore pressure drop $\Delta p_\mathrm{max}$ normalised by the undrained, isothermal pore pressure change $\Delta\Phi_\mathrm{max}/\beta^*$ as a function of hydraulic diffusion time relative to characteristic rupture time. Parameter values are $f_\mathrm{p}/f_\mathrm{r}=2$, $\delta_\mathrm{D}/\delta_\mathrm{w}=1$, $\chi=10$, $\Delta\Phi_\mathrm{max}/(\beta^*\sigma_0')=0.5$, and thermal pressurisation slip weakening distance is changed as reported on the graph.}
  \label{fig:pdrop}
\end{figure}

Is pore fluid vaporisation a widespread phenomenon during rupture? A simple assessment was made by \citet{brantut20} based on laboratory measurements of stick slip in granite, assuming isothermal and undrained conditions. Vaporisation was predicted to occur at modest slip distances (less than $1$~cm) under upper crustal conditions. This simple assessment can be revised by considering fluid diffusion and thermal pressurisation, which tend to limit the pressure drop associated to dilatancy. The maximum pore pressure drop is given by $\Delta\Phi_\mathrm{max}/\beta^*$, but a decreasing fraction of this maximum is achieved with increasing drainage and increasing contribution of thermal pressurisation (Figure \ref{fig:pdrop}). In the case of freshly fracture granite reported by \citet{brantut20}, the slip weakening distance associated to thermal pressurisation is most likely very large: the thermal pressurisation factor $\Lambda$ is small due to the large pore space compressibility of the newly formed gouge material, of the order of $10^{-8}$~Pa$^{-1}$. Combining this large compressibility with typical parameters for granite, using a fault thickness of $3$~mm, a friction coefficient of $0.6$, a heat capacity of $2.6$~MPa/K and a thermal pressurisation factor of $\Lambda=0.03$~MPa/K, we find $\delta_\mathrm{c}\approx0.4$~m. This is indeed much larger than the inferred frictional slip weakening distance and characteristic slip associated with dilatancy, which are of the order of $1$~mm. In that case, neglecting thermal pressurisation to compute pore pressure drop is justified, and the maximum pore pressure drop should remain close to the undrained limit $\Delta\Phi_\mathrm{max}/\beta^*$ if rupture is sufficiently fast (undrained tip). In laboratory experiments, direct measurements of pore pressure drop during faulting of intact granite are of the order of $30$~MPa for slip of the order of $1$~mm, so that vaporisation is expected in roughly the upper 3~km of the crust in such rock type \citep{brantut20,aben21}. With increasing temperature, the vaporisation pressure increases up to the critical point, around $20$~MPa at $370^\circ$C, so that vaporisation could be expected down to $5$~km depth. Such estimates are necessarily coarse, and strong local variations are expected due the natural roughness of faults (e.g., in large pull-apart regions).

When thermal pressurisation is not efficient, as in the case of newly formed faults in granite, frictional heating is large, which can lead to melting of the fault material. The characteristic temperature rise for undrained, adiabatic behaviour in the presence of rapid dilatancy at the onset of slip is given by $(\sigma'_0 + \Delta\Phi_\mathrm{max}/\beta^*)/\Lambda$ (see Appendix \ref{ax:TP_au}). At very large slip distances compared to $\delta_\mathrm{c}$, the effect of dilatancy becomes negligible (see Appendix \ref{ax:TP_sp}). A thorough examination of the onset of melting during thermal pressurisation was given by \citet{rempel06}; their analysis remains valid at large slip, and only needs to be amended at small slip by accounting for an initial offset in pore pressure equal to $\Delta\Phi_\mathrm{max}/\beta^*$.

In the case of granite, the slip at which melting occurs is considerably shortened by dilatancy. For instance, at a depth of 5~km, assuming lithostatic normal stress and hydrostatic initial pore pressure, the initial effective stress is $\sigma'_0=90$~MPa. With an average thermal pressurisation factor of $\Lambda=0.1$~MPa/K, intermediate between the low end-member value of $0.03$ reported in \citet{brantut20} for fresh fractures and a high end-member value of $0.3$~MPa/K reported in \citet{brantut16b} for mature granite gouge, the adiabatic, undrained temperature rise is of $900$~K. Melting is therefore expected only at slip distances several times larger than $\delta_\mathrm{c}$, (here a few centimetres if we assume a shear zone width of a few millimetres), well into the regime where strength is already significantly reduced by thermal pressurisation. With a pressure drop due to dilatancy of the order of $30$~MPa, commensurate to that measured in the laboratory, the adiabatic, undrained temperature rise becomes $1200$~K, so that only a fraction of $\delta_\mathrm{c}$ is required to trigger bulk melting of the shear zone, i.e., before any significant strength reduction due to thermal pressurisation.

Thus, dilatancy dramatically facilitates frictional melting. The common occurrence of pseudotachylytes in low-porosity, consolidated rocks \citep{sibson06} is therefore expected not only because of the potential ``dryness'' of the initial material, but because dilatancy would either vaporise any preexisting fluid or trigger fast melting due to the induced pore pressure drop \citep[see also][where this is discussed with an example from the field]{brantut18c}.

\subsection{Diffusion-driven rupture nucleation and growth}

As observed in dynamic simulations (see illustrative example in Figure \ref{fig:dilexample1}), dilatancy has a dramatic impact on the dynamics of rupture during the nucleation phase. A thorough analysis of the transition from aseismic to seismic slip in slip-weakening, dilatant faults has been conducted by \citet{ciardo19} in the context of injection-induced slip. Their analysis only considered along-fault fluid diffusion, assuming impermeable fault walls. Here, we considered the alternative hypothesis whereby only across-fault fluid flow was accounted for, which is reasonable when the pore pressure gradient along the $x$ direction (which scales with $1/L_\mathrm{w}$) is much smaller than that along the fault-perpendicular direction (which scales with $1/h$). Despite this difference, one important result of \citeauthor{ciardo19}'s analysis remains valid, and follows from the stress intensity factor computation given in Equation \eqref{eq:Kdyn}: there exists a minimum dilatancy amplitude above which rupture stabilisation can occur, given by
\begin{linenomath}
\begin{equation} \Delta\Phi_\mathrm{max}^\mathrm{c}=\beta^*\sigma'_0(\tau_\mathrm{b}/\tau_\mathrm{r}-1).
\end{equation}
\end{linenomath}
At $\Delta\Phi_\mathrm{max}>\Delta\Phi_\mathrm{max}^\mathrm{c}$, the undrained (isothermal) residual strength is above the background stress and rupture cannot propagate unless some fluid flows back into the fault zone. The rate of fault growth is then entirely determined by the rate of pore pressure recharge. The dynamic simulation shown in Figure \ref{fig:dilexample1} corresponds to the case $\Delta\Phi_\mathrm{max}=\Delta\Phi_\mathrm{max}^\mathrm{c}$, and a exhibits a large nucleation timescale compared to typical elastodynamic timescales. In addition to the stabilising effect due to offsets in residual strength in undrained conditions, dilatancy also has the potential to produce transient slip-strengthening behaviour at the crack tip, which considerably increases the nucleation size of ruptures loaded with relatively uniform stress \citep{brantut15}.

While a thorough analysis considering two-dimensional hydraulic diffusion and finite crack dimensions should eventually be conducted to make quantitative estimates of critical nucleation dimension and time, it is clear that dilatancy has a first order impact on fault stability and rupture initiation. Low porosity rocks, either intact, healed or sealed (e.g., granites) are the most affected by this process.


\subsection{Dynamics of slow ruptures driven by dilatancy}

In models based on rate and state friction laws, dilatancy hardening has been shown to be a viable mechanism for the generation and propagation of slow slip events in subduction zones \citep{segall10}. There are many characteristics of slow slip events that need to be explained quantitatively by any proposed physical model: spatio-temporal dynamics of events, recurrence rate, moment-duration relationship, relationship with earthquake location and timing, etc \citep[e.g.,][]{burgmann18}. Here, our very simple rupture model is not aimed at explaining all those observations; however, it is useful to test whether dilatancy hardening alone can explain the slowness of rupture propagation during slow slip events, and what parameter range would be required to do so.

Slow slip events in the Cascadia subduction zone typically propagate at speeds of the order of $10$~km/day. Their spatial extent ranges from $50$ to $100$~km, and they are associated to stress drops ranging from $2$ to $100$~kPa \citep{schmidt10,michel19}. The stress intensity factor at the tip of such ruptures can be approximated by \citet{hawthorne13, weng19}
\begin{linenomath}
\begin{equation}
  K=\psi\Delta\tau\sqrt{W},
\end{equation}
\end{linenomath}
where $\psi$ is a geometrical constant ($\approx 0.87$ in the geometry used by \citet{hawthorne13}, and given by $\approx \sqrt{2/\pi}$ for deeply buried antiplane faults according to \citep{weng19}), $\Delta\tau$ is the stress drop and $W$ is the width of the ruptured region. Using the spatial dimension and stress drop estimates from geodetic measurements, we obtain $K$ ranging from $0.5$ to $24$~MPa~m$^{1/2}$. This stress intensity factor has to be matched by the toughness that arises from frictional strength. Under purely drained conditions, the toughness is (Equation \ref{eq:Kcdry}) $K_\mathrm{c}^\mathrm{drained} = \sqrt{2\mu\delta_\mathrm{w}(f_\mathrm{p}-f_\mathrm{r})\sigma'_0}$. For slow slip events in subduction zones, we can reasonably assume that the fault zone material is not overconsolidated or well cemented, so that conventional rate and state friction is an appropriate description of the strength. Our simplified model of purely slip-weakening friction corresponds to the ``no-healing'' end-member of the slip law \citep{garagash21}; in that approximation, the friction drop at the rupture tip is therefore of the order of $f_\mathrm{p}-f_\mathrm{r}\sim10b$ \citep{hawthorne13,garagash21}, where $b$ is the evolution parameter of the friction law, itself of the order of $0.01$. The slip weakening distance is of the order of $1$ to $100$~$\mu$m. The effective stress in the regions experiencing slow slip events is low, of the order of a few megapascals, as inferred by seismological observations \citep[][Section 2.3]{burgmann18}. Using a shear modulus of $\mu=30$~GPa, we arrive at $K_\mathrm{c}^\mathrm{drained}$ ranging from $0.1$ to $1.7$~MPa~m$^{1/2}$, which overlaps with the $K$ estimate only when very low stress drops are considered in conjunction with large slip weakening distances and effective stress. In addition, rupture events are not expected to be stable if we consider only a balance between $K$ and $K_\mathrm{c}^\mathrm{drained}$, since any increase in $K$ would lead to dynamic acceleration of the rupture tip \citep{weng21}. When dilatancy is accounted for, ruptures are stabilised since any increase in $K$ and rupture speed can be matched by a corresponding increase in $K_\mathrm{c}$ (Figure \ref{fig:Kc}; see below).

As rule of thumb, for any stabilisation effect to be significant, i.e., for $K_\mathrm{c}$ to increase significantly above $K_\mathrm{c}^\mathrm{drained}$, the magnitude of the undrained pore pressure drop has to be (at least) comparable to that of the stress drop. If dilatancy is the stabilising process of slow slip events, we therefore expect the associated pore pressure drop to be, at the minimum, in the range $1$ to $100$~kPa.

\begin{figure}
  \centering
  \includegraphics{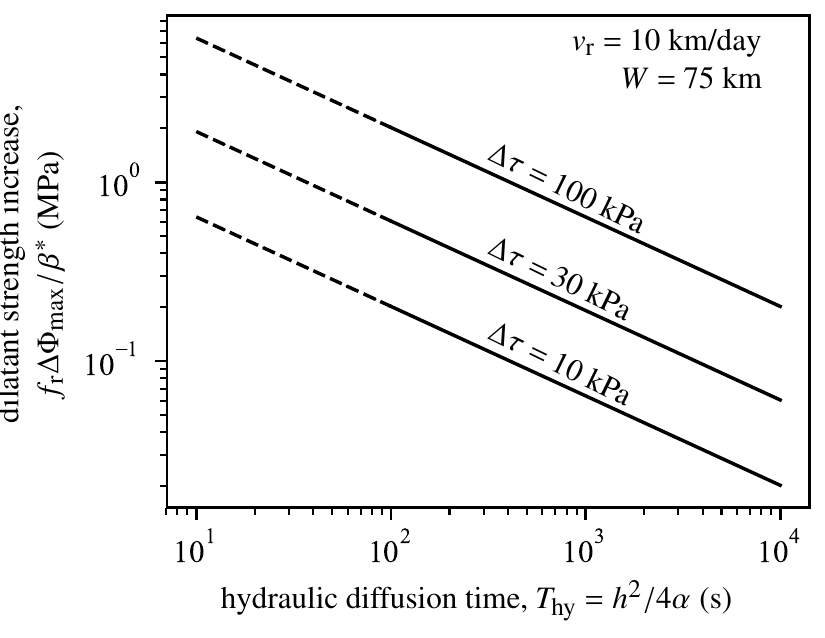}
  \caption{Combinations of dilatancy and diffusion parameters required to satisfy $K=K_\mathrm{c}$ for slow slip events for a range of stress drops. The frictional contribution to toughness is neglected compared to that of dilatancy. Dashed lines correspond to hydraulic diffusion times such that the characterstic distance $v_\mathrm{r}T_\mathrm{hy}$ is smaller than typical frictional cohesive zone size, rendering the use of Equation \eqref{eq:K_diff} imprecise.}
  \label{fig:slowslip}
\end{figure}

More precisely, the contribution of dilatancy and subsequent pore fluid diffusion can be estimated analytically using Equation \eqref{eq:K_diff}, assuming that the hydraulic diffusion time $T_\mathrm{hy}$ is large compared to the elastodynamic timescale $T=\mu\delta_\mathrm{w}/v_\mathrm{r}(f_\mathrm{p}-f_\mathrm{r})\sigma'_0$, that is, when the length scale $v_\mathrm{r}T_\mathrm{hy}$ is much larger than the frictional cohesive zone size. Choosing $\ell_\mathrm{cut}=10\times v_\mathrm{r}T_\mathrm{hy}$ (or, alternatively, $\ell_\mathrm{cut}=W$, with little quantitative impact on the result due to the weak dependency of $K_\mathrm{c}$ on the cuttoff length), we can look for the combination of parameters $T_\mathrm{hy}$ (hydraulic diffusion time) and $f_\mathrm{r}\Delta\Phi_\mathrm{max}/\beta^*$ (dilatant strength increase) required to match the stress intensity factor $K$ estimated from geophysical data. For stress drops of the order of $30$~kPa, rupture speeds of $10$~km/day can be explained by a combination of $f_\mathrm{r}\Delta\Phi_\mathrm{max}/\beta^*=0.3$ to $0.1$~MPa and $T_\mathrm{hy}=h^2/4\alpha_\mathrm{hy}=300$ to $3000$~s, respectively (Figure \ref{fig:slowslip}). This corresponds to an undrained pore pressure drop at the rupture tip of the order of $0.1$~MPa, so that only very modest dilatancy is required to severely limit the rupture speed, provided that diffusivity is low enough.

For a compressibility $\beta^*$ of the order of $10^{-10}$ to $10^{-9}$~Pa$^{-1}$, a pressure drop of $0.1$~MPa corresponds to a porosity change $\Delta\Phi_\mathrm{max}$ from $10^{-3}$ to $10^{-2}$\%, which is commensurate with the dilatancy parameter $\epsilon$ inferred by \citet{segall95} from \citet{marone90} data, and used by \citet{segall10} in their model for slow slip events. Even though the model used here is only slip-dependent and not directly rate-and-state dependent, both approaches point to relatively small dilatant effects (i.e., not of the order of tens of MPa as could be expected when fracturing intact material). Regarding pore fluid diffusion, in clay gouge, \citet{faulkner18} report diffusivities of the order of $10^{-8}$~m$^2$/s, so that shear zones of around $6$~mm width would be sufficiently thick to obtain diffusion time consistent with the model. If diffusivity of the fault zone material is comparable to that reported for exhumed continental faults, $\alpha\approx10^{-6}$~m$^2$/s \citep{wibberley03}, considerably thicker shear zones of $6$~cm would be needed to explain the rupture dynamics.

The parameter estimates obtained from our simple stress intensity factor analysis of slow slip events are realistic and consistent with laboratory measurements. As explained above, the rupture model used here is not designed to reproduce all the characteristics of slow slip events, but it does explain how slow rupture speeds can be achieved with dilatancy and pore fluid diffusion as the main control of strength.

Compared to the rate-and-state dependent approach used by \citet{segall10}, the slip-dependent friction and porosity model we have employed is considerably simpler, but leads to similar rupture dynamics. The computation of $K$ relies on the ruptures being confined in a strip of finite width \citep{hawthorne13}, which is not part of the assumptions used by \citet{segall10}. In a strictly two-dimensional configuration, the stress intensity factor is proportional to the squareroot of the rupture length. To see how ruptures might grow in this geometry, let us consider again, as in Section \ref{sec:energy}, a semi-infinite fault loaded by a background stress $\tau_\mathrm{b}>0$ in the region $x>0$, and $\tau_\mathrm{b}=0$ elsewhere. At a given time $t$, the rupture has grown to a distance $L$ in the loaded region, so that the stress intensity factor at the tip is
\begin{linenomath}
\begin{equation}
  K = 2(\tau_\mathrm{b} - \tau_\mathrm{r})\sqrt{\frac{2 L}{\pi}.}
\end{equation}
\end{linenomath}
In the small scale yielding approximation with cutoff length $\ell_\mathrm{c} =10\times v_\mathrm{r}T_\mathrm{hy}$, this stress intensity factor must equate the toughness given by Equation \eqref{eq:K_diff}, which is proportional to $\sqrt{v_\mathrm{r}} = \sqrt{dL/dt}$. Neglecting dynamic effects at small $v_\mathrm{r}$, we arrive at the following differential equation for $L(t)$:
\begin{linenomath}
\begin{equation}\label{eq:L_ode}
  \frac{1}{L}\frac{dL}{dt} = \frac{1}{T_\mathrm{hy}} \left[\frac{\tau_\mathrm{b}-\tau_\mathrm{r}}{f_\mathrm{r}\Delta\Phi_\mathrm{max}/\beta^*}\right]^2\mathrm{asinh}^{-2}\left(\sqrt{\pi\ell_\mathrm{c}/v_\mathrm{r}T_\mathrm{hy}}\right).
\end{equation}
\end{linenomath}
Assuming, as before, $\ell_\mathrm{c} =10\times v_\mathrm{r}T_\mathrm{hy}$ (keeping in mind that the dependency on $\ell_\mathrm{cut}$ is very weak), the last term in \eqref{eq:L_ode} is a constant, approximately equal to $0.17$. The solution of \eqref{eq:L_ode} is thus an exponential growth:
\begin{linenomath}
\begin{equation}
  L(t) = L(0)e^{t/t_\mathrm{g}},
\end{equation}
\end{linenomath}
where the growth rate is
\begin{linenomath}
\begin{equation}
  t_\mathrm{g} \sim 5.9\times T_\mathrm{hy}\left[\frac{f_\mathrm{r}\Delta\Phi_\mathrm{max}/\beta^*}{\tau_\mathrm{b}-\tau_\mathrm{r}}\right]^2.
\end{equation}
\end{linenomath}
Using the parameter values discussed above, with a stress drop of the order of $\tau_\mathrm{b}-\tau_\mathrm{r} = 0.01$~MPa, a dilatant strength contribution of the order of $f_\mathrm{r}\Delta\Phi_\mathrm{max}/\beta^*=0.1$~MPa, and a diffusion time of $T_\mathrm{hy}=10^3$~s, we obtain a growth rate of $t_\mathrm{g}\sim 5.9\times10^{5}$~s, i.e., about one week. Such a timescale is commensurate to the total duration of slow slip events \citep[e.g.,][]{michel19}, and is much longer than any timescale associated with wave propagation along the fault. While the exponential growth might be viewed as unstable, the small growth rate, essentially governed by the diffusion timescale, makes these ruptures far less sensitive to variations in stress (or, energy release rate) compared to ruptures driven by constant toughness (or, fracture energy). In that sense, dilatancy produces relatively stable ruptures, which is consistent with observations of slow slip events in subduction zones.




\section{Conclusions}


By analysing a crack model incorporating slip weakening friction and slip-dependent dilatancy coupled to fluid flow, we have established that dilatancy tends to limit crack propagation by increasing the stress intensity factor required for crack growth. At sufficiently large rupture speeds compared to the rate of fluid flow, the rupture tip is undrained, and the frictional strength is larger than in the nondilatant case. Further away from the tip fluid flow (i.e., pore pressure recharge) leads to a progressive decrease in strength back to the drained residual strength.

In a small-scale yielding approximation, the effect of dilatancy amounts to an increase in toughness that scales with the squareroot of rupture speed. This approximation is valid when the rupture dimension is much larger than the along-fault hydraulic diffusion length scale. For shorter rupture dimensions, the strength variations due to pore fluid diffusion are not confined in a small near-tip region, and the approximation of small-scale yielding becomes less accurate. Nevertheless, dilatancy still has the effect of increasing the applied stress intensity factor necessary for crack growth, which can be decomposed in a near-tip contribution (a relative decrease in toughness compared to the nondilatant regime) and a nonlocal contribution (a reduction in the net stress intensity factor) which includes a transition from undrained strength near the tip to the drained strength away from the tip. This natural decomposition of stress intensity factor into identifiable contributions does not have a simple parallel in terms of energy release rate and fracture energy, since $G$ depends quadratically on $K$, which couples near-tip and ``far''-tip effects.

When effective, thermal pressurisation is seen to compensate somewhat the strengthening due to dilatancy through an increase in heating rate. The resulting temperature increase is transiently larger than in nondilatant cases, which can facilitate frictional melting. However, for large slip distances, the effect of dilation on temperature and strength vanishes due to thermo-hydraulic diffusion with the surrounding medium.

The dilatancy toughening effects analysed here have direct applications in the study of the dynamics of slow slip events. A stress intensity factor analysis shows that the slowness of ruptures propagating in low effective stress regimes and with overall low stress drops, as in the case of slow slip events documented in the Cascadia subduction zone, can be explained using a simple fracture propagation model with a set of parameter values consistent with laboratory data. The relative stability of ruptures governed by slip weakening and dilatancy is linked to typically slow rupture growth rates, of the order of weeks, which is primarily controlled by the long hydraulic diffusion timescales inferred for slow slip events.




\paragraph{Acknowledgments}   Discussions with Frans Aben, Dmitry Garagash, Jessica Hawthorne, Jim Rice and Rob Viesca helped shape this work. Useful suggestions and comments from Huihui Weng and an anonymous reviewer contributed to improve the manuscript. Support from the UK Natural Environment Research Council (grants NE/K009656/1 and NE/S000852/1) and from the European Research Council under the European Union's Horizon 2020 research and innovation programme (project RockDEaF, grant agreement \#804685), is gratefully acknowledged. The results of this paper can be reproduced by direct implementation of the analytical formulae and numerical methods described in the main text and appendices; no new data were generated in this work.

\appendix

\section{Elastodynamic simulations}
\label{ax:dyn}

The elastodynamic equilibrium equation for 2D, mode III ruptures is
\begin{linenomath}
\begin{equation}\label{eq:dyn}
  \tau(x,t) = \tau_\mathrm{b}(x) - \frac{\mu}{2 c_\mathrm{s}} V(x,t) + \phi(x,t),
\end{equation}
\end{linenomath}
where $\phi(x,t)$ corresponds to static and dynamic contributions to stress associated with a given slip rate history on the fault. The stress in \eqref{eq:dyn} must be less than (if slip rate is zero) or equal to the fault strength (if slip rate is nonzero), given by Equation \eqref{eq:tauf}, together with governing equations \eqref{eq:f}, \eqref{eq:p(t)}, and \eqref{eq:Dphi}.

Numerical solutions for stress and slip rate are determined by the method outlined in \citet{brantut19b}, and only the general principle is recalled here. Time and space are discretised with constant steps. At each time step, the previous slip rate history is used to compute the stress functional $\phi$; the spatial convolution involved in this computation is performed in the Fourier domain. Then, the strength is evaluated by computing the pore pressure distribution along the fault, approximating the convolution integral in \eqref{eq:p(t)} by the mid-point rule. An estimate of slip rate at the next time step is then found by equating strength with stress. The procedure is then repeated, using the average between this estimate of slip rate and that from the previous time step in the computation of the stress transfer functional and pore pressure, to arrive at the final estimate of slip rate at the next time step. The evolution of friction coefficient and porosity in space and time is tracked by updating slip at each time and space step.

Space was discretised in $4096$ steps of size $\Delta x/L_\mathrm{w}' = 1/10$, where we recall that $L_\mathrm{w}'=\mu\delta_\mathrm{w}/\tau_\mathrm{r}$, and time was discretised in $8192$ steps of size $\Delta t = (1/2)\Delta x/c_\mathrm{s}$. In the simulation illustrated in Figure \ref{fig:dilexample1}, a uniform background stress equal to $\tau_\mathrm{b} = 1.5\tau_\mathrm{r}$ was used, and rupture was nucleated by setting $\tau_\mathrm{b} = 1.01(f_\mathrm{p}/f_\mathrm{r})\tau_\mathrm{r}$ in a region of size $6L_\mathrm{w}'$ around $x=0$.

\section{Rupture tip trajectory from LEFM}
\label{ax:eofm}

The energy balance is given by
\begin{linenomath}
  \begin{equation}
    G(v_\mathrm{r}) = G_\mathrm{c}^\mathrm{undrained},
  \end{equation}
\end{linenomath}
where the fracture energy is constant (Equation \eqref{eq:Gcundrained}). The dynamic energy release rate is a function of rupture speed, and can be approximated by \citep[e.g.][]{ampuero06}
\begin{linenomath}
  \begin{equation}
    G(v_\mathrm{r}) \approx \sqrt{\frac{1-v_\mathrm{r}/c_\mathrm{s}}{1+v_\mathrm{r}/c_\mathrm{s}}} \frac{K_\mathrm{static}^2}{2\mu},
  \end{equation}
\end{linenomath}
where $K_\mathrm{static}$ is the static stress intensity factor for the finite rupture in consideration, function of the current rupture tip positions and stress drop. Here we only consider symmetric ruptures, and denote the rupture tip position $a(t)$, so that the rupture speed is $v_\mathrm{r}=da/dt$. The static stress intensity factor is
\begin{linenomath}
  \begin{equation} \label{eq:kstat}
    K_\mathrm{static}(t, a) = \sqrt{\pi a}\frac{2}{\pi}\int_{-a}^{+a}\frac{\Delta\tau(x,t)}{\sqrt{a^2-x^2}}dx,
  \end{equation}
\end{linenomath}
where the stress drop is
\begin{linenomath}
  \begin{equation}
    \Delta\tau(x,t) = \tau_\mathrm{b}(x) - \tau_\mathrm{f}(x,t).
  \end{equation}
\end{linenomath}
The frictional strength $\tau_\mathrm{f}$ varies along the crack due to time-dependent diffusion of pore fluid after the passage of the rupture tip. All the near-tip slip-dependent, undrained contribution to strength has been accounted for in the undrained fracture energy and is thus not involved in the stress drop. At a given time $t$ and position $x$ ($|x|<|a|$), the strength is approximated by the large-time limit (Equation \eqref{eq:tau_diff})
\begin{linenomath}
  \begin{equation} \label{eq:tauf_diff_tr}
    \tau_\mathrm{f}(x,t) = \tau_\mathrm{r} \frac{\Delta\Phi_\mathrm{max}/(\beta^*\sigma'_0)}{\sqrt{1 + (t-t_\mathrm{r}(x))/T_\mathrm{hy}}},
  \end{equation}
\end{linenomath}
where $t_\mathrm{r}(x)$ denotes the time at which the rupture tip was at position $x$, i.e., $t_\mathrm{r}$ is the solution of $a(t)=x$.

The energy balance yields a delay differential equation for the rupture tip position,
\begin{linenomath}
  \begin{equation}
    \frac{da}{dt} = \frac{1 - (G_\mathrm{c}^\mathrm{undrained}/K_\mathrm{static}(t,a)^2)^2}{1 + (G_\mathrm{c}^\mathrm{undrained}/K_\mathrm{static}(t,a)^2)^2},
  \end{equation}
\end{linenomath}
which is solved numerically using the Julia \texttt{DifferentialEquations.jl} package \citep{rackauckas17}. The integral in \eqref{eq:kstat} is evaluated numerically using Gauss-Kronrod quadrature, and the time $t_\mathrm{r}(x)$ is obtained by root finding (bisection method). The initial condition for $a$ is set to $18\times L'_\mathrm{w}$, at the point of initial arrest immediately outside the nucleation zone.

\section{Thermal pressurisation and dilatancy}

Here some closed-form solutions are presented for the thermal pressurisation and dilatancy problem, with the assumption that the coefficient of friction remains constant. In that case, the solutions are sufficiently simple to allow clear physical interpretations.

\subsection{Adiabatic, undrained solution}
\label{ax:TP_au}

Neglecting fluid and heat diffusion, the governing equations for pore pressure and temperature simplify into two ordinary differential equations for effective stress and temperature:
\begin{linenomath}
\begin{align}
  \displaystyle \frac{d\sigma'}{d\delta} & = -\frac{\sigma'}{\delta_\mathrm{c}} + \frac{\Delta\Phi_\mathrm{max}}{\beta^*}\frac{e^{-\delta/\delta_\mathrm{D}}}{\delta_\mathrm{D}},\\
  \displaystyle \frac{d\Theta}{d\delta} & = \frac{\sigma'}{\Lambda\delta_\mathrm{c}}.
\end{align}
\end{linenomath}
Straightforward integration leads to the following solution:
\begin{linenomath}
\begin{equation}
  \sigma'(\delta) = \sigma'_0 e^{-\delta/\delta_\mathrm{c}} + \frac{\Delta\Phi_\mathrm{max}/\beta^*}{\delta_\mathrm{D}/\delta_\mathrm{c}}\big(e^{-\delta/\delta_\mathrm{D}}-e^{-\delta/\delta_\mathrm{c}}\big)
\end{equation}
\end{linenomath}
for effective stress and
\begin{linenomath}
\begin{equation}
  \Theta(\delta) = \Theta_0 + \frac{1}{\Lambda}\left[\sigma'_0(1-e^{-\delta/\delta_\mathrm{c}}) + \frac{\Delta\Phi_\mathrm{max}/\beta^*}{\delta_\mathrm{D}/\delta_\mathrm{c}}\left(\frac{\delta_\mathrm{D}}{\delta_\mathrm{c}}(e^{-\delta/\delta_\mathrm{D}})-(1-e^{-\delta/\delta_\mathrm{c}})\right) \right]
\end{equation}
\end{linenomath}
for temperature. It is readily observed that the temperature asymptotically increases by a quantity $(\sigma'_0+\Delta\Phi_\mathrm{max}/\beta^*)/\Lambda$ at large slip, so that dilatancy has the effect of resetting the initial effective stress, as reported by \citet{rice06}.

\subsection{Slip on a plane, constant slip rate solution}
\label{ax:TP_sp}

For Gaussian shear strain rate profiles and dilatancy proportional to strain rate, the pore pressure evolution is given in integral form by
\begin{linenomath}
\begin{align}
  p(0,t)-p_0 &= \frac{\Lambda}{\rho c}\int_0^t\tau(t') V\left[\frac{\alpha_\mathrm{th}/(\alpha_\mathrm{th}-\alpha_\mathrm{hy})}{\sqrt{h^2+4\pi\alpha_\mathrm{th}(t-t')}}-\frac{\alpha_\mathrm{hy}/(\alpha_\mathrm{th}-\alpha_\mathrm{hy})}{\sqrt{h^2+4\pi\alpha_\mathrm{hy}(t-t')}}\right]dt'\nonumber\\
  &\qquad\qquad -\frac{1}{\beta^*}\int_0^t\frac{N(t')V}{\sqrt{h^2+4\pi\alpha_\mathrm{hy}(t-t')}}dt',
\end{align}
\end{linenomath}
where $N$ is the rate of dilatant fault opening, given by $N/h = d\Delta\Phi/dt$. For vanishingly small $h$ compared to both thermal and hydraulic boundary layer widths, the expression simplifies to
\begin{linenomath}
\begin{equation}
  p(0,t)-p_0 = \frac{\Lambda}{2\rho c (\sqrt{\alpha_\mathrm{th}}+\sqrt{\alpha_\mathrm{hy}})}\int_0^t \frac{\tau(t')}{\sqrt{\pi(t-t')}}dt' - \frac{1}{2\beta^*\sqrt{\alpha_\mathrm{hy}}}\int_0^t \frac{N(t')}{\sqrt{\pi(t-t')}}dt'.
\end{equation}
\end{linenomath}
Further assuming constant friction and slip rate, we arrive at the following integral form for shear stress:
\begin{linenomath}
\begin{equation}
  \tau(\delta)-\tau_0 = -\frac{1}{\sqrt{L^*}}\int_0^\delta \frac{\tau(\delta')}{\sqrt{\pi(\delta-\delta')}}d\delta' + \frac{1}{\sqrt{D}}\int_0^\delta \frac{N(\delta')/\beta^*}{\sqrt{\pi(\delta-\delta')}}d\delta',
\end{equation}
\end{linenomath}
where $L^*=4(\sqrt{\alpha_\mathrm{hy}}+\sqrt{\alpha_\mathrm{th}})^2(\rho c/f_\mathrm{r}\Lambda)^2/V$ and $D=4\alpha_\mathrm{hy}/f_\mathrm{r}^2V$.

The specific form \eqref{eq:Dphi} for dilatancy can be rewritten in terms of fault zone opening as
\begin{linenomath}
\begin{equation}
  N(\delta) = \Delta h/\delta_\mathrm{D}e^{-\delta/\delta_\mathrm{D}},
\end{equation}
\end{linenomath}
where $\Delta h$ is the maximum opening, equivalent to $\Delta\Phi_\mathrm{max}\times h$. The solution for shear stress can be found in the Laplace domain
\begin{linenomath}
\begin{equation}
  \hat{\tau}(s) = \tau_0 \frac{\sqrt{L^*/s}}{1+\sqrt{sL^*}} + \frac{\Delta h}{\delta_\mathrm{D}\beta^*}\sqrt{\frac{L^*}{D}}\frac{1}{(1+\sqrt{sL^*})(s+1/\delta_\mathrm{D})},
\end{equation}
\end{linenomath}
where $\hat{\tau}$ denotes the Laplace transform of shear stress and $s$ is the transformed independent variable. By standard inversion technique we find the following solution in terms of slip:
\begin{linenomath}
\begin{align}
  \tau(\delta) &= \tau_0\exp\left(\frac{\delta}{L^*}\right)\mathrm{erfc}\left(\sqrt{\frac{\delta}{L^*}}\right) + \frac{\Delta h}{\delta_\mathrm{D}\beta^*}\sqrt{\frac{L^*}{D}}\frac{1}{1+L^*/\delta_\mathrm{D}}\left[-\exp\left(\frac{\delta}{L^*}\right)\mathrm{erfc}\left(\sqrt{\frac{\delta}{L^*}}\right) +\right.\nonumber\\
  &\qquad\qquad \left. \exp\left(-\frac{\delta}{\delta_\mathrm{D}}\right)\left(1 + \sqrt{\frac{L^*}{\delta_\mathrm{D}}}\mathrm{erfi}\left(\sqrt{\frac{\delta}{\delta_\mathrm{D}}}\right)\right) \right].
\end{align}
\end{linenomath}
Reintroducing $\Delta\Phi_\mathrm{max}=\Delta h/h$, and denoting
\[
  \tau_\mathrm{D} = \frac{\Delta h}{\delta_\mathrm{D}\beta^*}\sqrt{\frac{L^*}{D}} = \frac{f_\mathrm{r}\Delta \Phi_\mathrm{max}}{\beta^*}\frac{\delta_\mathrm{c}}{\delta_\mathrm{D}}\left(1+\sqrt{\frac{\alpha_\mathrm{th}}{\alpha_\mathrm{hy}}}\right),
  \]
the shear stress can be rewritten with more easily recognisable quantities as
\begin{linenomath}
  \begin{align}
  \tau(\delta) &= \tau_0\exp\left(\frac{\delta}{L^*}\right)\mathrm{erfc}\left(\sqrt{\frac{\delta}{L^*}}\right) + \frac{\tau_\mathrm{D}}{1+L^*/\delta_\mathrm{D}}\left[-\exp\left(\frac{\delta}{L^*}\right)\mathrm{erfc}\left(\sqrt{\frac{\delta}{L^*}}\right) +\right.\nonumber\\
  &\qquad\qquad \left. \exp\left(-\frac{\delta}{\delta_\mathrm{D}}\right)\left(1 + \sqrt{\frac{L^*}{\delta_\mathrm{D}}}\mathrm{erfi}\left(\sqrt{\frac{\delta}{\delta_\mathrm{D}}}\right)\right) \right]. \label{eq:tau_sp}
\end{align}
\end{linenomath}
In the slip on a plane, constant slip rate assumption, the temperature evolution is given by
\begin{linenomath}
\begin{equation}
  \Theta(\delta) - \Theta_0 = \frac{1}{2\rho c}\sqrt{\frac{V}{\alpha_\mathrm{th}}}\int_0^\delta \frac{\tau(\delta')}{\sqrt{\pi(\delta-\delta')}}d\delta'.
\end{equation}
\end{linenomath}
Using again Laplace transforms and after a lengthy but straightforward
calculation, we arrive at
\begin{linenomath}
\begin{equation}\label{eq:T_sp}
  \Theta(\delta)-\Theta_0 = \frac{1}{f_\mathrm{r}\Lambda}\left(1+\sqrt{\frac{\alpha_\mathrm{hy}}{\alpha_\mathrm{th}}}\right)\left[\tau_0-\tau(\delta) + \tau_\mathrm{D}\sqrt{\frac{\delta_\mathrm{D}}{L^*}}e^{-\delta/\delta_\mathrm{D}}\mathrm{erfi}\left(\sqrt{\frac{\delta}{\delta_\mathrm{D}}}\right)\right].
\end{equation}
\end{linenomath}

The main interest of closed form expressions \eqref{eq:tau_sp} and \eqref{eq:T_sp} is that they show that (1) shear stress approaches zero and (2) temperature approaches $\sigma'_0(1+\sqrt{\alpha_\mathrm{hy}/\alpha_\mathrm{th}})/\Lambda$ at large slip. The latter asymptote is the same as that without dilatancy \citep{rice06}. The effect of dilatancy is thus \emph{not} simply to reset the initial pore pressure: since dilatancy only occurs within the fault, pore pressure will spontaneously reequilibrate with the surrounding medium, so that at large time the effect of dilatancy becomes insignificant.


\end{document}